%% file: paper.tex
\documentclass[]{fairmeta}

\usepackage{makecell}

\addto\extrasenglish{

}

\title{Collective Communication for 100k+ GPUs}

\author[]{Min Si}
\author[]{Pavan Balaji}
\author[]{Yongzhou Chen}
\author[]{Ching-Hsiang Chu}
\author[]{Adi Gangidi}
\author[]{Saif Hasan}
\author[]{Subodh Iyengar}
\author[]{Dan Johnson}
\author[]{Bingzhe Liu}
\author[]{Regina Ren}
\author[*]{Deep Shah}
\author[]{Ashmitha Jeevaraj Shetty}
\author[*]{Greg Steinbrecher}
\author[]{Yulun Wang}
\author[]{Bruce Wu}
\author[]{Xinfeng Xie}
\author[]{Jingyi Yang}
\author[]{Mingran Yang}
\author[*]{Kenny Yu}
\author[]{Minlan Yu}
\author[]{Cen Zhao}
\author[]{Wes Bland}
\author[]{Denis Boyda}
\author[]{Suman Gumudavelli}
\author[]{Prashanth Kannan}
\author[]{Cristian Lumezanu}
\author[]{Rui Miao}
\author[]{Zhe Qu}
\author[]{Venkat Ramesh}
\author[]{Maxim Samoylov}
\author[]{Jan Seidel}
\author[]{Srikanth Sundaresan}
\author[]{Feng Tian}
\author[]{Qiye Tan}
\author[]{Shuqiang Zhang}
\author[]{Yimeng Zhao}
\author[]{Shengbao Zheng}
\author[]{Art Zhu}
\author[]{James Hongyi Zeng}
\contribution[*]{Work done at Meta}

\input{abstract}

\correspondence{James Hongyi Zeng}
\date{\today}

\metadata[Code]{\url{https://github.com/meta-pytorch/torchcomms/tree/main/comms/ncclx}}

\begin{document}

\maketitle

\input{intro}
\input{background}
\input{stack_overview}
\input{training}

\input{inference}
\input{operation}
\input{related}
\input{conclusion}
\input{acknowledgements}

\clearpage
\newpage
\bibliographystyle{assets/plainnat}
\bibliography{paper}

\end{document}

%% file: abstract.tex
\abstract{
The increasing scale of large language models (LLMs) requires highly efficient GPU communication mechanisms, particularly as training workloads extend to hundreds of thousands of GPUs. Traditional communication methods face significant throughput and latency limitations at this scale, hindering both the development and deployment of state-of-the-art models. This paper presents the NCCLX collective communication framework, developed at Meta and  engineered to optimize performance across the full LLM lifecycle, from the synchronous demands of large-scale training to the low-latency requirements of inference. The framework is designed to support complex workloads on clusters exceeding 100,000 GPUs, ensuring reliable, high-throughput, and low-latency data exchange. Empirical evaluation on the Llama4 model demonstrates substantial improvements in communication efficiency. This research contributes a robust solution for enabling the next generation of LLMs to operate at unprecedented scales.
}

%% file: intro.tex
\section{Introduction}
The rapid advancement of large language models (LLMs) has driven the need for highly efficient distributed machine learning systems. Collective communication between GPUs is fundamental to scaling performance. As model sizes and training datasets continue to expand, the challenge of efficiently coordinating hundreds of thousands of GPUs across clusters and data centers becomes increasingly complex. Ensuring reliable, high-throughput, and low-latency communication is essential for both the development and deployment of these models.

To support the development and deployment of the Llama4 model, we developed NCCLX, a collective communication framework based on the popular NCCL~\citep{nccl} library, with the following features: 

{\bf Scalability and performance:} NCCLX supports collective communications for over 100K GPUs while meeting the throughput and latency requirements for training and inference applications.  In training, massive clusters of GPUs operate synchronously to train the initial model, often reaching scales of 100k+ GPUs for state-of-the-art workloads. Specifically for large scale training beyond thousands of GPUs, NCCLX enables fast initialization and fault tolerant communication to support high ratio of effective training. In inference, the trained model is deployed for real-world applications, requiring low-latency communication to serve user requests efficiently in distributed serving environments.

{\bf Support custom features for diverse model applications:} There are diverse set of models for training and inference applications, with different parallelism schemes, communication collectives, and customized features. NCCLX offers flexible APIs and collective primitives to support these diverse model requirements with high performance.  

{\bf Easy to program:}  Our communication framework should make it straightforward for developers to program, customize, and optimize across the entire communication stack. NCCLX allows developers to easily plug in new algorithms, transport backends, and optimization solutions to meet the needs of diverse model applications.

{\bf Operational tools for monitoring, debugging, and maintenance:} NCCLX provides robust tooling for monitoring, debugging, and maintaining large-scale deployments of collective communications. 

\begin{table}[t]
\centering
\begin{tabular}{ | m{1.7cm} | m{7.4cm} | m{6cm} | }
\hline
 \textbf{Phase} & \textbf{Unique problems and requirements} & \textbf{Features} \\ \hline 

  Generic & Leverage host-driven framework to support diverse model communication requirements and custom algorithms & Host-driven customization (\autoref{sec:host-driven-customization}) \\ \hline

 Generic & SM-free communication to avoid interference in compute/communication overlap scheme especially in multi-dimensional parallelism & Zero-copy data transfer (\autoref{sec:zero-copy-transfer}) \\ \hline

 Generic & Co-design with PyTorch's memory management for user buffer registration under implicit lifetime management in zero-copy data transfer & Tensor registration management for zero-copy (\autoref{sec:tensor_registration_management}) \\ \hline

 Generic & Avoiding network congestion when handling concurrent large volume network traffic, while keeping high sending speed &  CTran and network co-design (\autoref{sec:dqplb}) \\ \hline

 Training & Enable low-latency communication over extended paths while optimizing GPU resource consumption in Pipeline Parallelism & Zero-copy and SM-free send/receive (\autoref{sec:pp-zero-copy-send-recv}) \\ \hline

 Training & Achieve fine-grained communication and computation overlap in Tensor Parallelism & RMA Put (\autoref{sec:tp-rma-put}) \\ \hline
  
 Training & Achieve no-hang operation, elasticity, and flexible restart in Hybrid Sharding Data Parallel & Fault tolerant AllReduce (\autoref{sec:ftar}) \\ \hline
 
 Multi-node Inference  & Avoid any data padding in computation imbalance scheme (e.g., MoE) with CUDA graph & GPU-resident collectives (\autoref{sec:gpu_resident_collectives}) \\ \hline

 Multi-node Inference & Reduce CPU overhead for small messages & Low-latency optimizations (\autoref{sec:low_latency_optimization}) \\ \hline

 Control Plane & Scalable initialization among training hosts before starting communication on high-speed RoCE/IB network & Scalable initialization in training (\autoref{sec:scalable_initialization}) \\ \hline

 Tooling & Keep high communication performance with minimal GPU memory, GPU SM, and network resources (QPs, etc.) & Resource management (\autoref{sec:resource}) \\ \hline
 
 Tooling & Fast faulty hardware localization; fast user fault debugging in multi-dimensional parallelism programming & Fault localization (\autoref{sec:fault_analyzer})  \\ \hline
\end{tabular}
\caption{Key features and the target workloads and problems.}
\label{tab:problems}
\end{table}

NCCLX operates beneath the PyTorch \citep{torchcomms} layer and manages all communications for both training and inference processes. 
NCCLX provides users a rich selection of communication semantics in three execution modes: Host-initiated APIs, Host-initiated APIs with GPU-resident metadata, and Device-initiated APIs. Each of the execution models further provides collectives, point-to-point, and remote memory access (RMA) semantics. 

To support these communication semantics, we developed a host-driven custom transport layer called CTran. CTran supports a wide variety of communication algorithms with various topology-based optimizations, zero-copy and SM-free transfers, and custom features (e.g., fault tolerance). CTran contains NVLink, Infiniband/RoCE (IB) and socket backends to support lower-level communication primitives via different hardware routines with unique load balancing optimizations for the IB/RoCE backend.

With NCCLX, we observed performance gains in both training and inference scenarios compared to NCCL. During training, NCCLX reduced the latency of each steady training step of Llama4 models by up to $12\%$ across various scales. The scalable initialization represents up to $11\times$ faster training startup time at 96K scale. For inference with the Llama4 Maverick model, NCCLX achieved end-to-end decoding latency improvements ranging from $15\%$ to $80\%$ across diverse distributed configurations.

\autoref{tab:problems} summarizes the key features introduced by NCCLX for Llama4 workloads. The primary contributions include host-driven customization, zero-copy and SM-free communication, which jointly support diverse model communication requirements and custom algorithms while avoiding interference with compute/communication overlap schemes. Building on these core features, we further introduce specialized customizations for both training and multi-node inference scenarios.

For the rest of the paper, we will first cover the network and existing vendor communication stacks as a background. We then give the overview of NCCLX and CTran, and discuss their use cases in large-scale training and inference. We end the paper with evaluation and some future works.

%% file: background.tex
\section{Background}
In this section, we will walk through the cluster's hardware setup, as well as the communication library basics.

\subsection{ML Training and inference frameworks}
\textbf{Training.} Training large LLMs involves multi-dimensional model parallelism, which generates multiple concurrent medium-to-large network traffic flows with diverse performance characteristics. Collectives in inner parallel domains (e.g., tensor parallelism, TP) are often fully exposed in training execution and remain within high-bandwidth domains. Middle parallel domains, such as expert parallelism (EP) and pipeline parallelism (PP), introduce partially hidden communication with medium message sizes, but exposed overhead may be serialized (e.g., send/receive in PP). In the outermost domains (e.g., fully sharded data parallelism, FSDP, and data parallelism, DP), collectives carry large data volumes but are often well hidden by inner domains.

These massive jobs must be scheduled across GPU clusters and even across data center buildings, resulting in multi-layer hierarchical network topologies with varying network latency and switch congestion tolerance at different network layers. 

\textbf{Inference.}
Unlike training, inference involves low-latency communication for small to medium-size messages. Inference computation can be parallelized across multiple GPUs and nodes without communication bottlenecks. MoE AllToAll is a well-known collective at LLM inference, with a few MB of data transfer per operation. While network transfers between GPU pairs can be overlapped, CPU instructions to prepare network requests are serialized and can take more time than the data transfer itself. The community has demonstrated the efficiency of NVSHMEM in MoE AllToAll scenarios, thanks to its optimized low-overhead instructions and ability to leverage multiple GPU threads for parallelizing instructions to independent peers~\citep{deepep}.

\subsection{Communication libraries}

Driven by the demands of Deep Learning and emerging LLM applications, GPU vendors provide native communication libraries alongside their hardware. For example, NVIDIA offers NCCL~\citep{nccl} and NVSHMEM~\citep{nvshmem}, each designed for different communication patterns. AMD follows a similar trend, so we will not discuss their offerings in detail here.

NCCL employs a host-initiated communication model, where the CPU schedules communication and defines input arguments as CPU variables. It is commonly used for GPU communication in traditional DL/ML applications, where bulk synchronization is prevalent. Collectives like AllReduce, AllGather and ReduceScatter, are commonly used in various model domains (e.g., Data Parallelism, Tensor Parallelism). Since these collectives typically involve medium to large data volumes, NCCL prioritizes bandwidth utilization and data flow entropy over latency. Additionally, NCCL's execution model is designed for regular collective patterns where communication arguments (e.g., size, data type) can be statically expressed as host arguments. While this model is well-suited for traditional DL/ML scenarios and simplifies collective definition for model developers, it lacks flexibility for more dynamic communication, such as when communication arguments are generated by preceding computations and vary. Passing such arguments to an NCCL collective necessitates copying values from GPU to host memory. This not only introduces additional CPU synchronization, potentially delaying subsequent kernel scheduling, but also proves incompatible with CUDA graphs, often requiring expensive data padding as a workaround.

In contrast, NVSHMEM utilizes device-initiated communication semantics. This means communication is invoked from the device kernel, and input arguments are defined as device variables. NVSHMEM is designed for low-latency and dynamic communication patterns. Data transfers can be initiated directly from within a GPU computation kernel and sent directly to the network, minimizing scheduling latency. Because communication is defined within the kernel function, arguments generated by a preceding kernel computation can be passed directly. These benefits, including low latency and kernel execution, make NVSHMEM ideal for developing fine-grained computation and communication pipelines. However, NVSHMEM has its limitations: It requires a symmetric memory region to be globally allocated and registered with the network on all GPU ranks. This occupied memory is not shared with PyTorch, so overuse of NVSHMEM can significantly restrict model scaling.

In real-world LLM applications, diverse communication patterns often coexist across different parallel domains. This frequently leads to a dual communication runtime scheme, where NVSHMEM handles specific domains, and NCCL serves the remainder of the application. Such a dual runtime introduces limitations in performance optimization and resource sharing, and it doubles the operational maintenance required for industry-scale workloads. To address these challenges, we propose supporting both communication semantics through a unified communication stack, which will be further detailed in the next section.

\subsection{The Network}\label{sec:background_network}

\begin{figure}[tp]
\centering
\includegraphics[width=.8\columnwidth]{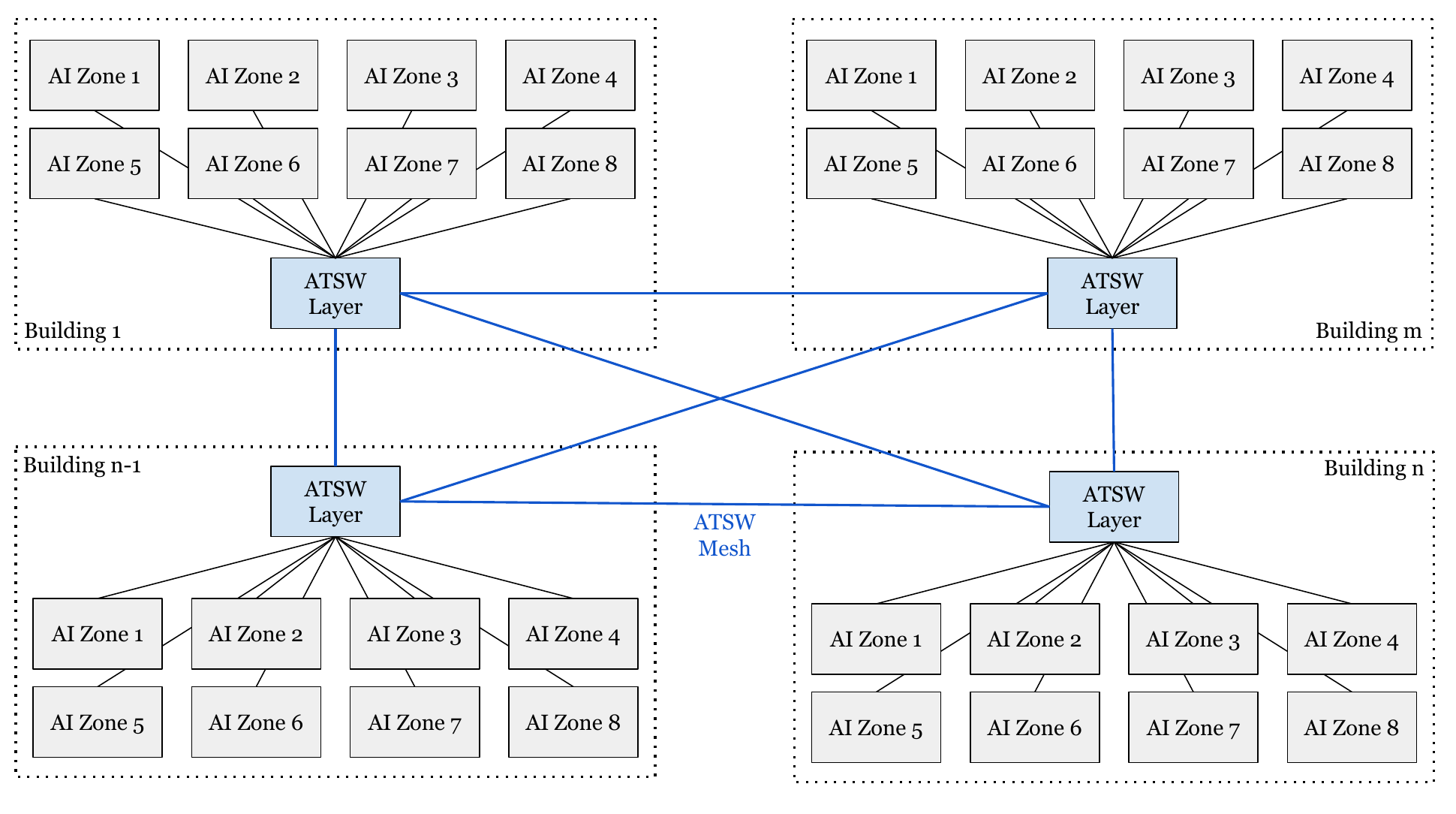}
\caption{Multi-building network architecture.}
\label{fig:network_architecture}
\end{figure}

A training cluster with over 100K GPUs often requires multiple datacenter (DC) buildings. To support this scale, we designed a multi-building network that is capable of integrating hundreds of thousands of GPUs across nearby DC buildings into a single high-performance RoCE fabric. We first describe the network architecture within a DC building, followed by the architecture across multiple DC buildings.

The network within a DC adopts a 3-layer Clos architecture, as shown in \autoref{fig:network_architecture}. Each DC is partitioned into multiple AI Zones. The Rack Training Switch (RTSW) connects GPUs within a rack, while the Cluster Training Switches (CTSW) connect all racks within an AI Zone. Aggregator Training Switches (ATSW) connect CTSWs across the DC, extending the RoCE network beyond a single AI Zone. Compared to the Llama 3 network, we reduced the cross-AI-Zone over-subscription ratio from 1:7 to 1:2.8 to provide higher bandwidth and better support multi-dimensional parallelism training at larger scale.

To interconnect multiple DCs, we use a fully connected mesh between the ATSW layers of different DC buildings as shown. Inter-DC traffic experiences the same over-subscription ratio as cross-AI- Zone traffic (1:2.8). This architecture is extensible and can scale to hundreds of thousands of GPUs within the same RoCE fabric, with DCs added incrementally over time. 

In a trainer Cluster connecting over 100K GPUs, GPU-to-GPU communication latency increases significantly with the number of network hops. Specifically, GPUs within the same rack have the lowest latency, while communication across different racks within the same AI zone, across AI zones, and across separate datacenter (DC) buildings experience 7$\times$, 15$\times$, and 30$\times$ higher latency, respectively. This increased latency is primarily due to cumulative switching delays and increased cable lengths. Achieving optimal collective performance at such scale and hierarchical latency requires the collective library to post messages that are large enough to saturate the bandwidth delay product (BDP), but not too much larger to create congestion with too much outstanding data. We talked about how we achieve this balance in  \autoref{sec:dqplb}.

More details of the cluster topology can be found in previous work \citep{gangidi2024RDMA}.

%% file: stack_overview.tex
\section{NCCLX Communication Stack Overview}

\begin{figure*}[tp]
\centering
\includegraphics[width=.8\columnwidth]{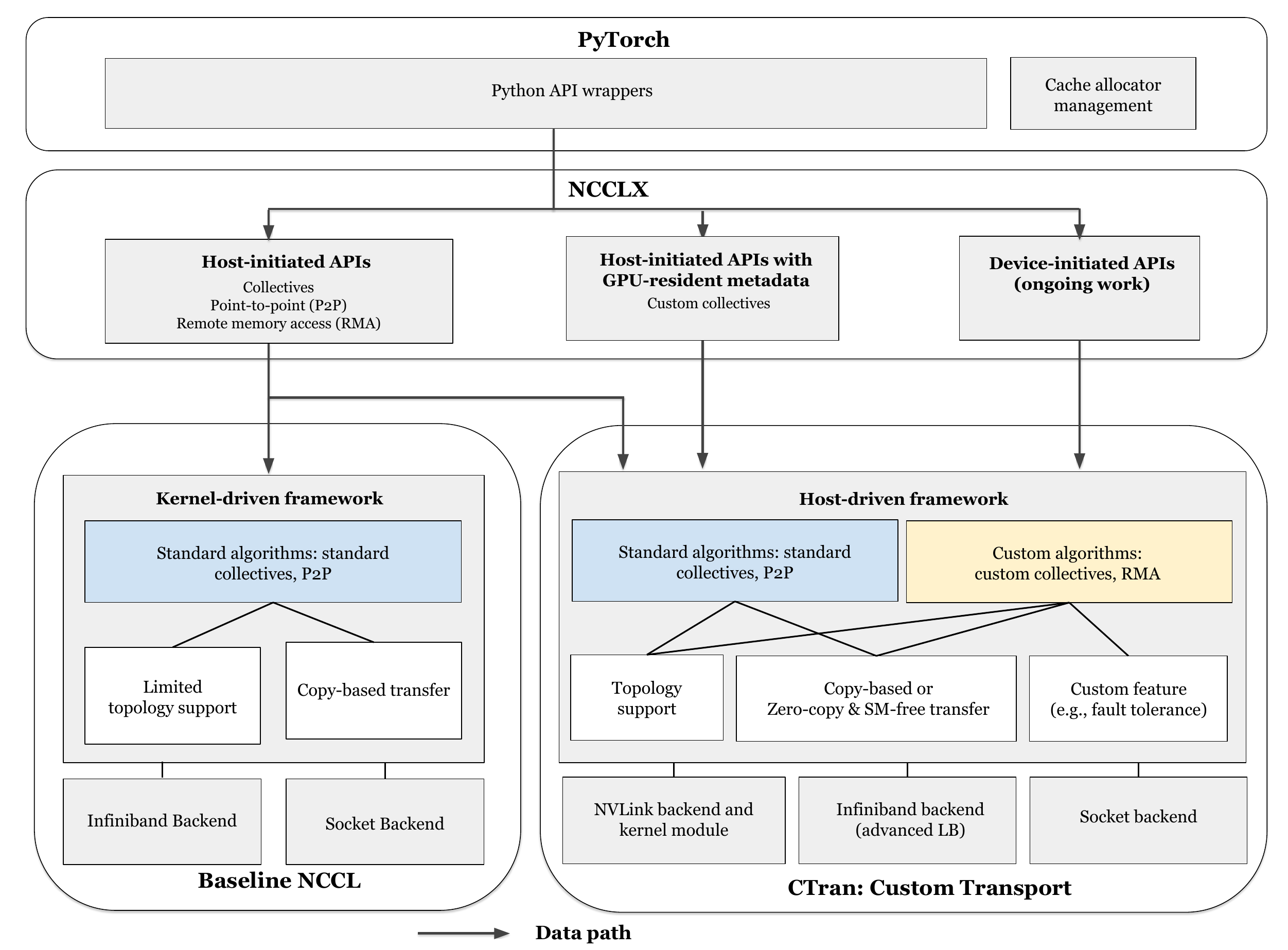}
\caption{NCCLX communication stack overview. NCCLX provides three categories of APIs: Host-initiated APIs, Host-initiated APIs with GPU-resident metadata, and Device-initiated APIs.
}
\label{fig:ncclx_stack_overview}
\end{figure*}

The key goal for NCCLX is to provide a high-performance, scalable, and customizable collective communication framework for diverse application model needs and diverse networking backends, while making it easier for communication developers to program. \autoref{fig:ncclx_stack_overview} gives an overview of the NCCLX stack.

NCCLX operates beneath the PyTorch layer and manages all the communications for both training and inference processes. 
NCCLX provides users a rich selection of communication semantics in three execution modes: Host-initiated APIs, Host-initiated APIs with GPU-resident metadata, and Device-initiated APIs. Each mode supports collective, point-to-point and remote memory access (RMA) semantics.

The fully host-initiated APIs removed unnecessary synchronizations between CPUs and GPUs, and provide higher performance than the baseline NCCL solution. The host-initiated APIs also enable developers to easily program custom collective algorithms such as send/recv in pipeline parallelism, AllGather in FSDP easier.

Sometimes, the decisions in collective algorithms are based on current metadata in the device, so fully host-initiated APIs are not enough. For example, in MoE training and inference systems, we may need the current routing information at the GPUs to adapt the collective algorithms. To handle such cases, we introduce the host-initiated APIs with GPU-resident metadata. 

As an ongoing effort, we are actively extending NCCLX to also provide device-initiated APIs. The device-initiated communication model is known to be beneficial for custom kernels that require intensive small messages and fine-grained compute-communication overlap, as widely demonstrated by NVSHMEM applications.~\citep{nvshmem} In this paper, we 
mainly focus on the underlying host-driven common communication stack and leave the rest of device-initiated model support as future work. 

To support the communication semantics, NCCLX develops a custom transport framework called CTran that follows a host-driven communication framework with the design principle to promote zero-copy and SM free communication when possible.
In addition to the key design differences, CTran further extends the baseline NCCL in four aspects: First, CTran supports all three execution models via the unified communication framework. Second, CTran supports a wide variety of communication algorithms including not only the standard collectives with topology-based optimizations, zero-copy P2P, but also custom collectives (e.g., fault tolerance) and remote memory access (RMA).  
Third, while providing all three of NVLink, Infiniband/RoCE (IB) and socket backends to support lower-level communication primitives via different hardware routines, we particularly provide an advanced load balancing solution for the RoCE backend. Finally, the communication critical path is highly optimized to minimize software overhead so that ensure low latency in small message scenarios.  We will detail the core design principle in \autoref{sec:ctran} and zoom into custom features and network load balancing via our training use cases in \autoref{sec:algorithm_performance_optimizations}. The host-initiated APIs with GPU-resident metadata mode and the low-latency optimizations will be deep dived in \autoref{sec:inference}, mainly driven by inference use cases.

When the model layer calls into a NCCLX communication, NCCLX dispatches the communication into either the baseline NCCL or the CTran code path. For custom communication operations (e.g., RMA, GPU-resident collectives) which don’t have an implementation in baseline NCCL, they are directly dispatched to CTran. For classical NCCL collectives and point-to-point, we allow users to explicitly choose the underlying baseline or CTran algorithms via environment variables. When deploying to a ready-to-launch model, we often partner NCCLX with offline auto-tuning, so that the optimal algorithms can be automatically selected.

NCCLX also provides scalable initilization and monitoring and diagnosis tools, which we will talk more in \autoref{sec:other}.

%% file: training.tex
\input{training/zero_copy}

\input{training/zero_copy_eva}

\input{training/algos}
\input{training/algos_eva}

%% file: training/zero_copy.tex
\section{CTran: The Custom transport in NCCLX}
\label{sec:ctran}

Our initial exploration of Llama4-scale training using NVIDIA NCCL revealed two limitations: a kernel-driven design and copy-based data transfers. To address these fundamental limitations, we developed a custom transport communication stack—CTran—based on zero-copy and SM-free communication, and a host-driven algorithmic framework.

\subsection{Host-driven customization}
\label{sec:host-driven-customization}

\begin{figure}[tp]
    \centering
    \includegraphics[width=\linewidth]{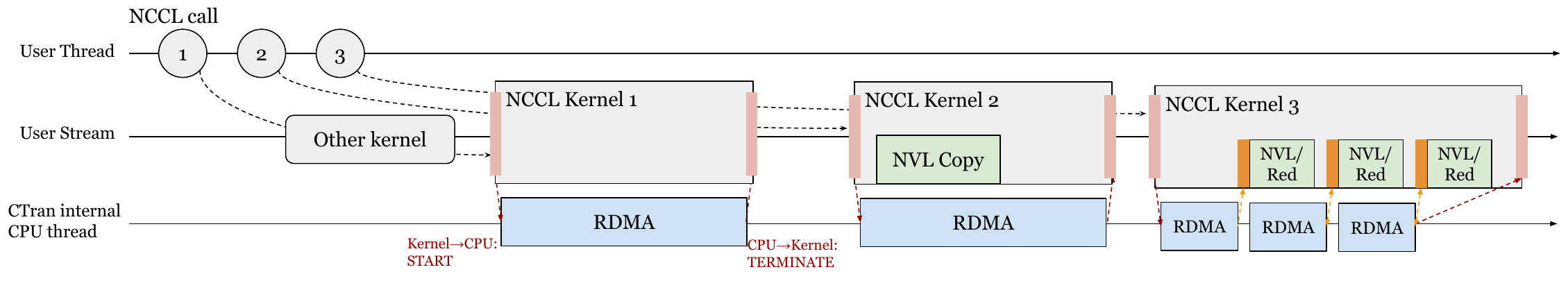}
    \caption{Coordination between CTran internal CPU thread and the CUDA kernel for a NCCL collective. Local D2D copy (i.e., for out-of-place collective) and P2P copy to NVLink peers are handled within ncclKernel, while RDMA network transfer is driven by the CPU thread. NCCL Kernel 1 demonstrates the fully-host-driven mode; NCCL Kernels 2 and 3 demonstrate the host-kernel coordinated mode.}
    \label{fig:host_driven}
\end{figure}

One of the main goals of NCCLX is to make it easy to support diverse model communication requirements and for communication developers to program custom algorithms. 
Unlike NCCL, where collective algorithms are mostly executed within CUDA kernels and internal RDMA operations are scheduled by host proxy threads, the CTran stack follows a host-driven framework for host-initiated collectives. Specifically, CTran launches a dedicated CPU background thread for each communicator. When a user program invokes a NCCL collective, CTran schedules the collective algorithm on the CPU thread while launching a stall kernel on the user-specified stream. The stall kernel ensures the communication follows the stream ordering semantics. Synchronization between the stall kernel and the CPU-side algorithm is achieved using a lightweight, host-pinned flag at both the start and end of the communication. \autoref{fig:host_driven} illustrates this coordination.

CTran enables us to rapidly deploy classical collective algorithms for large-scale training that can hide the communication time within the training pipeline. Furthermore, we were able to co-design custom communication routines with model algorithms, including fine-grained compute-communication pipelines for tensor parallelism (TP).

For collectives that involve only network data transfer (e.g., inter-node only AllGather in Fully Sharded Data Parallelism, point-to-point in Pipeline Parallelism), each RDMA operation can be directly posted from the CPU thread without any synchronization needed with the kernel side, significantly reducing the latency of small and medium message size collectives. We categorize it as \textbf{fully-host-driven mode}. NCCL Kernel 1 in \autoref{fig:host_driven} demonstrates this mode.

For collectives that require both network and NVLink data transfer, we extend the scheduling framework to support a \textbf{host-kernel coordinated mode}. For instance, the kernel would directly perform NVLink copy in parallel with CPU side RDMA if there is no dependency between the two operations. AllToAll in Mixture of Experts  falls into such a pattern (see NCCL Kernel 2 in \autoref{fig:host_driven}). For algorithms with certain operation dependency (e.g., inter-node AllReduce in Data Parallelism where in-kernel reduce and network RDMA are pipelined; detailed in \autoref{sec:ftar}), the kernel and CPU thread synchronize via lightweight producer-comsumer flags allocated from host-pinned memory. NCCL Kernel 3 in \autoref{fig:host_driven} illustrates this mode.  We have measured the synchronization overhead is always less than a microsecond and can be hidden even in complex pipeline algorithms. We will detail the overhead analysis in the following subsection.

\subsection{Zero-copy data transfer}
\label{sec:zero-copy-transfer}

\begin{figure}[tp]
\begin{subfigure}[b]{0.55\textwidth}
\centering
\includegraphics[height=0.17\textwidth]{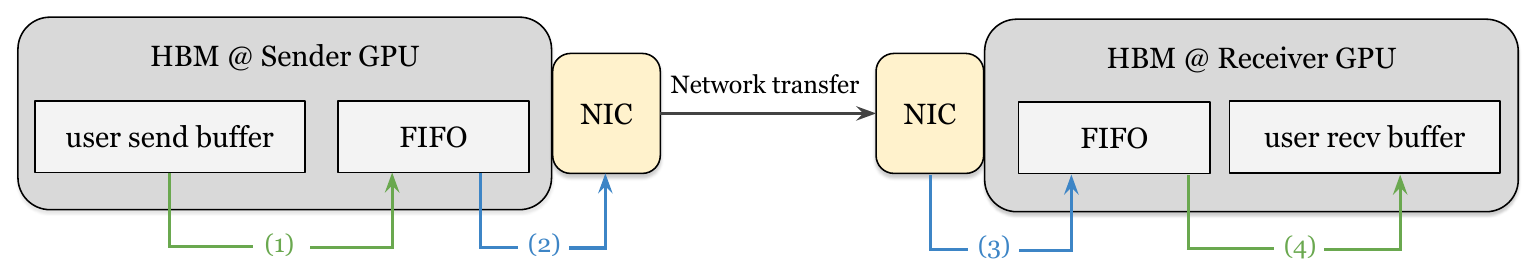}
\caption{Copy-based transfer.}
\label{fig:copy_breakdown}
\end{subfigure}
\begin{subfigure}[b]{0.33\textwidth}
\centering
\includegraphics[height=0.3\textwidth]{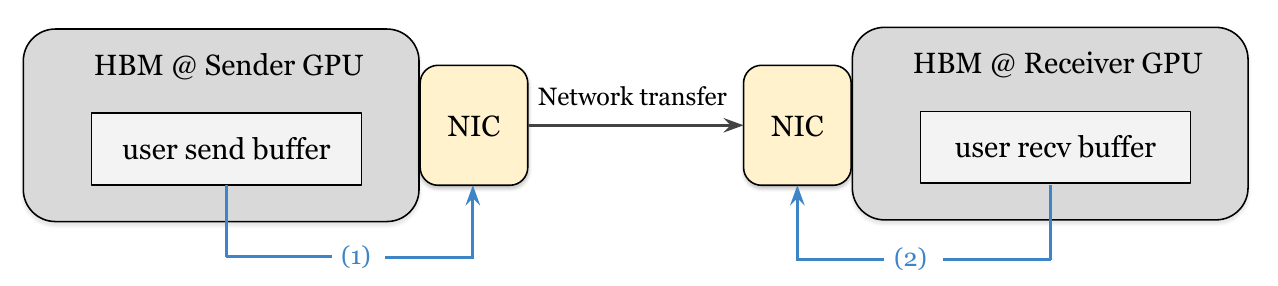}
\caption{Zero-copy transfer.}
\label{fig:zerocopy_breakdown}
\end{subfigure}
\caption{Data transfer breakdown from user send buffer to receive buffer via network transfer. Internal FIFO buffer is required in copy-based transfer. In addition to the required network transfer from sender side NIC to receiver side NIC, (a) copy-based transfer includes: (1) D2D copy from send buffer to sender side FIFO, (2) PCIe transfer from sender side FIFO to NIC, (3) PCIe transfer from NIC to receiver side FIFO, and (4) D2D copy from receiver side FIFO to receive buffer . In contrast, (b) includes: (1) PCIe transfer from send buffer to sender side NIC, and (2) PCIe transfer from receiver side NIC to receiver buffer.}
\end{figure}

\autoref{fig:copy_breakdown} illustrates the copy-based data transfer mechanism used by baseline NCCL, which necessitates an additional device-to-device copy on both the sender and receiver sides. To initiate a transfer, the sender rank copies data from the user buffer into an NCCL-internal "FIFO buffer" that is pre-registered with the network. The data is then transmitted via RDMA from the sender’s FIFO buffer to the receiver’s corresponding FIFO buffer. Finally, the receiver rank copies the data from its FIFO buffer to the destination user buffer. A similar copy-based data transfer approach is used across GPUs within the NVLink domain. Note that the copy operations between the user buffer and the FIFO buffer (steps (1) and (4) in \autoref{fig:copy_breakdown}) are handled by the collective kernel’s Streaming Multiprocessors (SMs) and utilize High Bandwidth Memory (HBM) bandwidth. In contrast, PCIe transfers (steps (2) and (3)) are executed by the network adapter’s DMA engine, triggered by an RDMA request from an internal CPU proxy thread. As a result, these PCIe transfers do not require involvement from the GPU SMs. 

\begin{figure}[tp]
    \centering
    \includegraphics[width=1\linewidth]{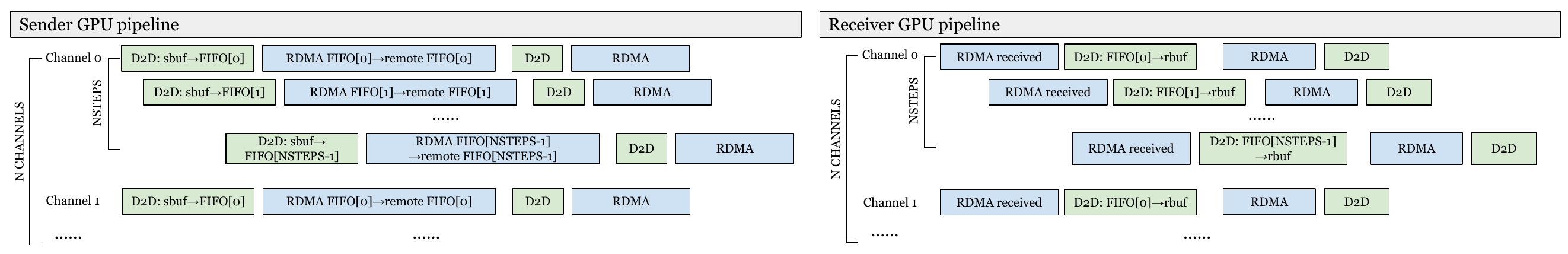}
    \caption{Copy + RDMA pipeline in copy-based transfer between two GPUs. Both sender and receiver employ \textit{NCHANNELS} number of pipelines, each called channel. Each channel is driven by a dedicated GPU thread block. Each channel further forms a \textit{NSTEPS}-way pipeline via two multi-slot FIFO buffers on sender and receiver, respectively. Sender side pipeline involves device-to-device copy from user send buffer (sbuf) to FIFO, followed by RDMA from the FIFO slot; receiver side pipeline starts with RDMA receive for a FIFO slot, followed by a device-to-device copy to the destination receive buffer (rbuf). %
    }
    \label{fig:copy_pipeline}
\end{figure}

Because in-device copy bandwidth is significantly higher than network or NVLink transfer speeds, the copy-based data transfer approach requires fine-grained data chunking and pipelining to overlap device-to-device copies with the slower network transfers. \autoref{fig:copy_pipeline} illustrates the copy-RDMA pipeline for a send-receive operation between two GPUs. This pipeline presents three limitations. First, the copy-based approach consumes GPU computing resources (SMs) and HBM bandwidth for copy operations between the user buffer and the FIFO buffer. This causes resource contention with concurrent computation and lead to performance degradation, forcing users to trade off resources between communication and computation. 
Second, this approach requires data to be segmented and transferred through multiple RDMA requests to establish the copy-RDMA pipeline. This process necessitates GPU-CPU synchronization at each pipeline stage and, more critically, limits each RDMA operation to a single data chunk. As a result, it restricts network utilization and makes it difficult to achieve full network saturation, particularly in high-latency environments such as cross AI-zone or cross data center (DC) building scenarios. Finally, each pipeline is managed by a dedicated thread block, known as a “Channel.” This static binding between pipelines and thread blocks requires allocating separate FIFO buffers for each independent pipeline. As a result, increasing the number of thread blocks to accelerate a collective operation leads to higher GPU memory consumption (see \autoref{sec:resource} for memory usage details).

To address these fundamental limitations, we develop the CTran stack with a zero-copy and SM-free communication design. Modern GPU systems, such as the NVIDIA H100~\footnote{Older architectures may lack full NVLink connectivity among all in-node GPUs, which is outside the scope of this paper.}, support direct data transfers between user buffers over both InfiniBand/RoCE networks and within the NVLink domain. With zero-copy, the entire data can be offloaded directly to the network transport layer, without the need for additional copies through internal FIFO buffers. \autoref{fig:zerocopy_breakdown} illustrates the data movement involved in the zero-copy transfer scheme.

This zero-copy design minimizes resource contention between concurrent communication and computation. For network-only collectives, we issue RDMA operations directly from the user source buffer to the destination buffer, eliminating the need for kernel involvement. For collectives within the NVLink domain, we utilize the CopyEngine where possible through custom user-facing APIs. While zero-copy is a well-established optimization in traditional HPC CPU communication, enabling efficient GPU buffer registration within the PyTorch ecosystem without incurring noticeable overhead presents unique challenges. We address these challenges to deliver a practical solution suitable for industry-scale production systems (see tensor registration management details in \autoref{sec:tensor_registration_management}).

The zero-copy design also facilitates flexible network configuration and optimization. With zero-copy, the entire data can be offloaded to the network transport layer without additional copies via internal buffers. The network transport layer can then decide to split the data into multiple RDMA packets and allocate to multiple queue pairs (QPs) for network traffic balancing. In this case, no pipelining is required at the algorithm layer
, and a CPU thread handles RDMA packet splitting by considering network saturation (see traffic load balancing details in \autoref{sec:dqplb}).

\subsection{CTran Design details}

\subsubsection{Tensor registration management for zero-copy}
\label{sec:tensor_registration_management}
For zero-copy operations, user buffers require registration for either NVLink or network direct transfer. This can be challenging given the implicit lifetime management within model applications. To address this, we co-designed our solution with PyTorch's memory management.

Specifically, network buffer registration typically requires the low-level network driver to identify the physical address ranges of both send and receive buffers so that network adapters can access them. This process usually involves physical page lookups and pin-downs at the OS kernel level, leading to registration times of hundreds of microseconds to a few milliseconds for the medium-sized buffers common LLM workloads. However, during GPU buffer registration in real-world scenarios, we've observed significant registration time spikes, occasionally extending to 100 milliseconds. Our investigation points to inter-process lock contention within the GPU RDMA driver as a primary cause. For example, memory allocation by other processes on the machine can delay a registration. We are currently collaborating with NVIDIA to identify the root cause and implement a fix.

By taking the potential overheads into account, we carefully extended the CUDA cache allocator (CCA) in Pytorch to provide two registration modes. The first mode we have deployed is \textbf{auto-registration}, where our pytorch  backend tracks and caches all CUDA segments allocated by the allocator~\footnote{Users can turn on the auto-registration by setting the environment variable \texttt{TORCH\_NCCL\_USE\_TENSOR\_REGISTER\_ALLOCATOR\_HOOK} to 1}. The ``tensor cache'' is managed within the CTran stack, by extending the ncclCommRegister API. CTran only caches these tracked tensor addresses when ncclCommRegister is called. The actual network registration of a tensor happens only when it is first time used in a collective. We called it ``lazy registration.'' For model stages that fall into a regular memory usage pattern without high memory pressure (e.g., AllGather in Fully Sharded Data
Parallelism which occurs at start of each training step), the auto-registration mode performs well.

While auto-registration mode is transparent to model-level programs, its effectiveness hinges on significant buffer reuse within CCA. Ideally, the same physical address ranges would be frequently reused for communication calls, making the initial buffer registration cost negligible. However, in practice, models often utilize CCA's expandable segment mode. This mode allows an allocated physical memory range (segment) to be remapped to different virtual address spaces to manage fragmentation. When this occurs, the previously registered buffer must be deregistered and the newly mapped virtual address space reregistered. In certain parallel domains, such as Pipeline Parallelism, frequent remapping can be triggered by high memory usage, leading to different physical memory ranges being used for repeated communication calls (e.g., send/receive). This results in a considerable registration overhead.

To mitigate frequent registration overhead, we implemented a \textbf{memory-pool mode}. This approach involves pre-allocating and registering a large memory pool from which all communication tensors are assigned. While this minimizes registrations, it requires explicitly labeling tensors for allocation from this separate pool. Although CUDA Graph could potentially automate this tensor relationship understanding and labeling, its deployment presents separate challenges and is not yet enabled in our pre-training workloads. A potential drawback of a dedicated memory pool is its impact on memory efficiency, as pre-allocated pools can reduce the maximum memory available for compute tensors. To mitigate this, we enhanced PyTorch CCA, enabling the default pool to utilize free space from pre-allocated pools when nearing out-of-memory (OOM) conditions. \footnote{Users can turn memory reuse from MemPool by setting \texttt{use\_on\_oom} to true at MemPool creation.~\citep{pytorch-mempool}}

\subsubsection{Diverse collective algorithms at host CPUs}

AllGather and ReduceScatter collectives in the Data Parallel (DP) domain are heavily challenged at large scale workloads due to the high network latency at cross-CTSW and cross-zone network domain (see network topology in \autoref{sec:background_network}). NCCL only had the Ring algorithm till the recent 2.23 release introduced the PAT algorithm~\citep{nccl-pat}. To unblock our early stage large scale training before PAT is available, we ported the latency-optimized Brucks and Recursive Doubling algorithms for AllGather, Recursive Vector-Halving Distance-doubling algorithm for ReduceScatter, and Tree algorithm for Broadcast, respectively~\citep{mpialgo}. These algorithms have been widely used in the classical High Performance Computing domain for large scale CPU systems. Thanks to the host-driven framework of CTran, it was straightforward to port the classical CPU communication algorithms. We omit the details in this paper.

\subsection{CTran and network co-design}
\label{sec:dqplb}

At scales exceeding 100K GPUs, GPU-to-GPU communication latency increases significantly with the number of network hops. Specifically, GPUs within the same rack have the lowest latency, while communication across different racks within the same AI zone, across AI zones, and across separate datacenter (DC) buildings experience 7$\times$, 15$\times$, and 30$\times$ higher latency, respectively. This increased latency is primarily due to cumulative switching delays and increased cable lengths. 

The two-stage copy mechanism within baseline NCCL further amplifies the impact of network latency. In this mechanism, clear-to-send control messages from the receiver to the sender are placed on the critical path. Consequently, when latency-sensitive collectives from the innermost layers of parallelism traverse cross-DC links, training performance suffers significant degradation. Even data-parallel collectives—which are generally more tolerant of network latency—are adversely affected when using baseline NCCL. To overcome these limitations, we introduce several optimizations in this section.

As described in \autoref{sec:ctran}, to mitigate the high control-message latency of the two-stage copy solution, we implemented zero-copy collectives. These enable the NIC to perform RDMA directly between source and destination buffers, eliminating the need for an intermediate buffer in the collective library and avoiding extra GPU-driven copies. As a result, GPU-to-GPU exchanges require only a single control message from receiver to sender, significantly reducing network latency. 

One drawback of the zero-copy communication is that the entire message is handed off to the network hardware at once, relying solely on the network fabric for both flow control and congestion control. This approach is not well-suited to our congestion control strategy. As noted in prior work~\citep{gangidi2024RDMA}, we do not employ traditional congestion control mechanisms such as DCQCN to limit switch buffer occupancy. Instead, we rely on deep-buffer switches to absorb transient bursts and leverage receiver-driven flow control in the collective library to prevent persistent congestion. However, zero-copy communication reduces opportunities for receiver feedback, increasing the likelihood of posting larger messages at once, which potentially leads to network overwhelming and excessive buffer build-up. Our evaluation confirms this analysis: using zero-copy communication alone resulted in suboptimal performance. In contrast, copy-based communication inherently segments data into smaller chunks—due to temporary buffer size constraints—thereby providing implicit flow control. 

To combine the advantages of both approaches, we employ zero-copy communication while internally partitioning data into smaller message segments and rate-limiting the number of segments in flight. To this end, we develop Dynamic Queue Pair Load Balancing (DQPLB) technique, which is a design within CTran to configure the amount of outstanding data on a per-connection and per-topology basis, allowing higher limits for cross-DC or cross-AI-Zone links, where the bandwidth-delay product (BDP) is greater than links within an AI Zone or rack. This fine-grained control enables more effective management of buffer build-up. Together with network load balancing improvements in prior work~\citep{gangidi2024RDMA} and spine switch Virtual Output Queuing (VOQ) tuning, compared to Llama3 training, we reduce switch buffer build-up by an order of magnitude in the RoCE network used in Llama4 training.

\subsubsection{DQPLB Design}
\begin{figure}
\begin{subfigure}[b]{0.59\textwidth}
\centering
\includegraphics[width=\textwidth]{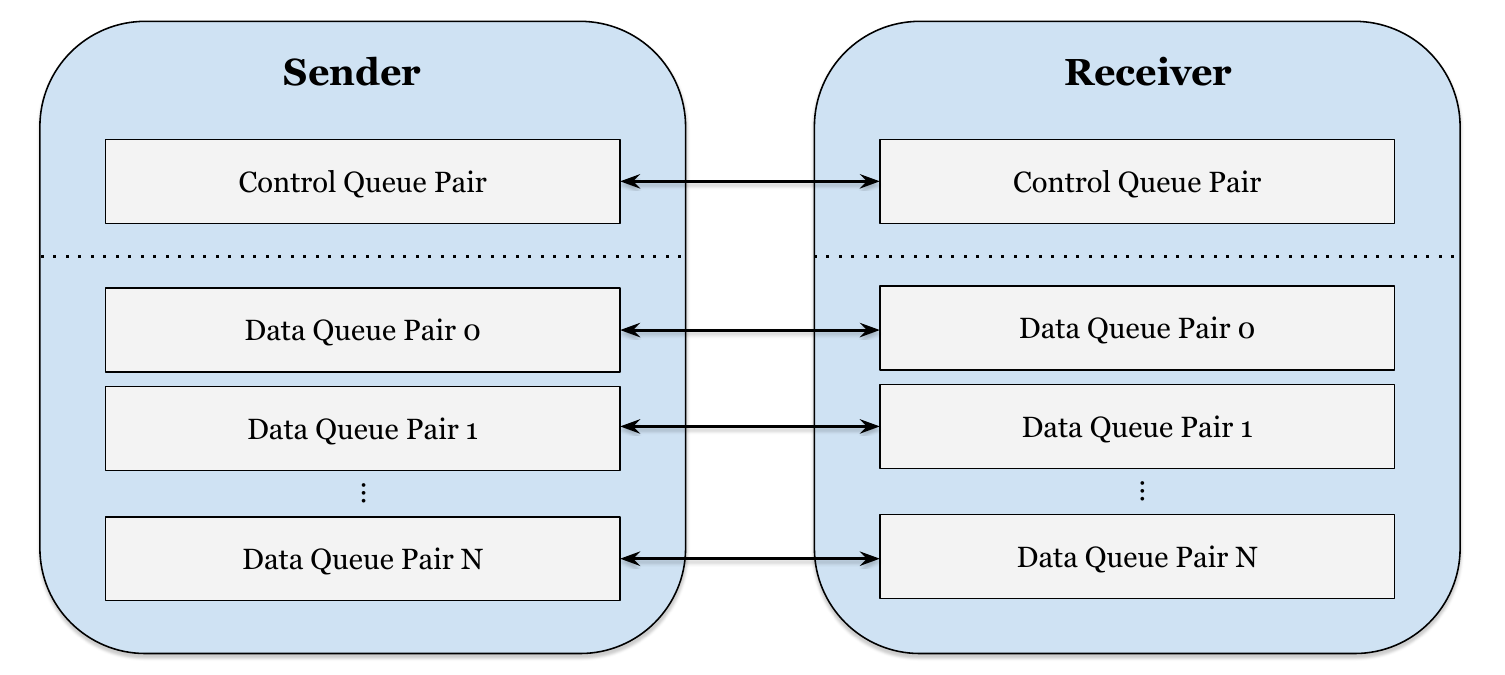}
\caption{Queue pair framework.}
\label{fig:ctran_dqplb}
\end{subfigure}
\begin{subfigure}[b]{0.41\textwidth}
\centering
\includegraphics[width=0.95\textwidth]{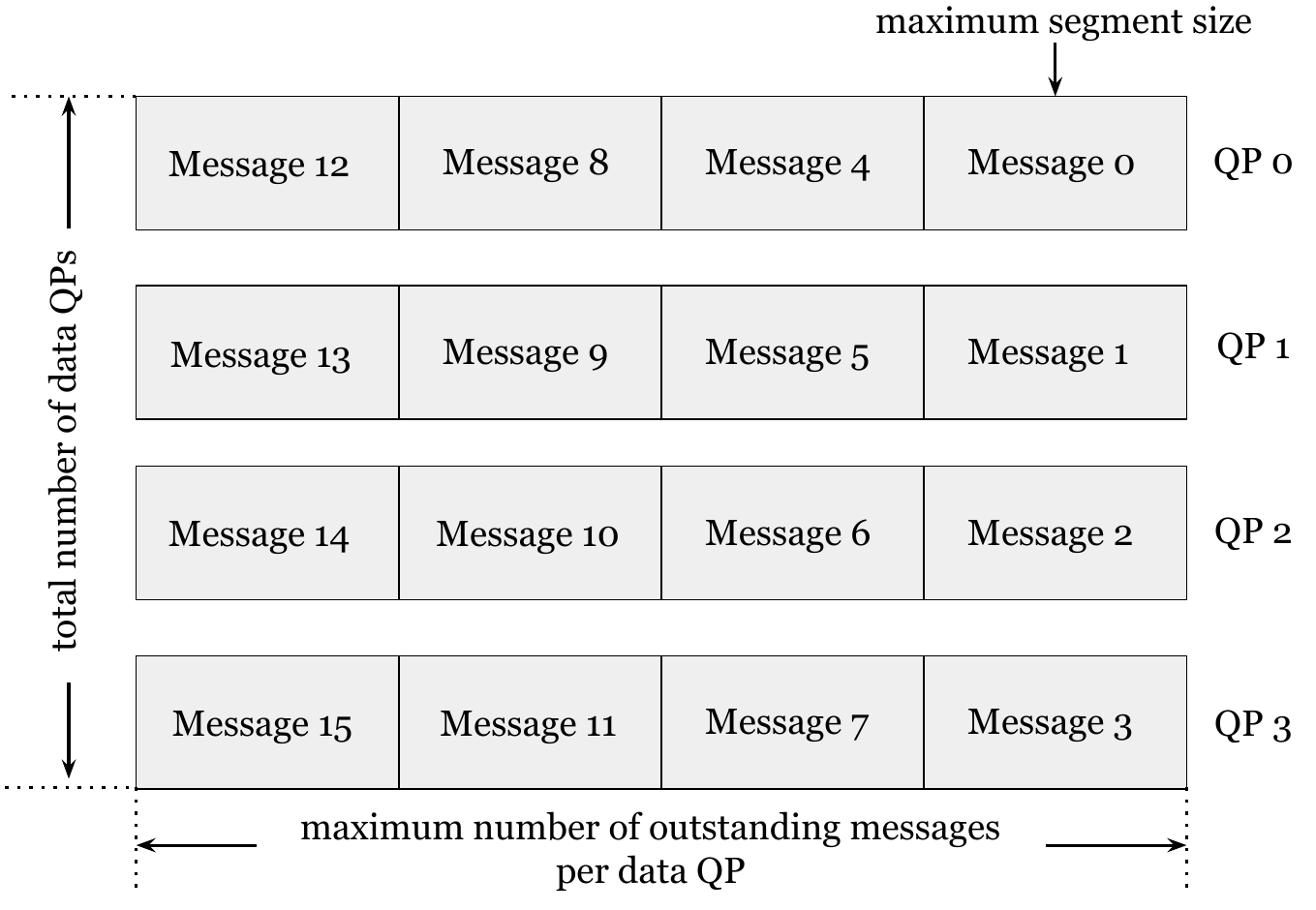}
\caption{Example of load balancing across four QPs.}
\label{fig:ctran_dqplb_lb}
\end{subfigure}
\caption{DQPLB design overview.}
\end{figure}

As shown in \autoref{fig:ctran_dqplb}, DQPLB uses one control queue pair (QP) and one or more data QPs. The control QP is responsible for exchanging memory addresses at the start of a collective operation, while the data QPs handle data transmission over the scale-out network, which forms the core of the DQPLB load balancing mechanism.

The number of data QPs is configurable and can be adjusted based on network topology and performance requirements. Our observations indicate that, with zero-copy communication, allowing an endpoint to send unlimited data does not always yield optimal performance. Instead, limiting the total amount of outstanding data per connection helps reduce congestion in the fabric and speeds up collective completion. Accordingly, as shown in~\autoref{fig:ctran_dqplb_lb}, we impose limits on the total number of data QPs, the maximum number of outstanding (unacknowledged) messages per data QP, and the maximum segment size, all on a per-connection-type basis.

We define four categories of connection types: within the same rack, cross-rack within the same zone, cross-zone within the same DC, and cross-DC. Based on the relative proximity of GPUs, we adjust the outstanding message limits for each connection. For connections with closer proximity, we use more conservative settings to accommodate the lower BDP. In contrast, for distant connections, we employ more aggressive configurations—such as higher number of data QPs and higher number of maximum number of outstanding messages—to better utilize the higher network BDP.

\subsubsection{DQPLB Operation}

DQPLB achieves ordered message delivery across multiple data QPs by leveraging sequence numbering, immediate data encoding, and out-of-order message tracking. The algorithm distributes messages to data QPs in a round-robin fashion. When a data QP’s pending work queue reaches its configured limit, message transmission is paused until a completion queue element (CQE) is received on the corresponding completion queue (CQ). Upon receiving a CQE, the system resumes posting work queue elements (WQEs) for that QP. This dynamic allocation enables queue pairs with lower network contention to transmit data more efficiently, allowing them to handle a greater traffic load.

To ensure ordered message delivery, we maintain two sequence counters: the sender tracks the next sequence number to transmit, while the receiver tracks the next expected sequence number. When transmitting messages in DQPLB mode, the sender utilizes the \texttt{IBV\_WR\_RDMA\_WRITE\_WITH\_IMM} opcode for InfiniBand post send operations, instead of the standard \texttt{IBV\_WR\_RDMA\_WRITE}. This enables the embedding of control information within the 32-bit immediate data field: bits 0–23 encode the sequential message number, bit 30 indicates fast path usage (explained in detail later), and bit 31 serves as a notification flag for the final write of a multi-packet message. When a message exceeds the maximum segment size, it is divided into multiple WQEs, with each partition assigned a consecutive sequence number; only the final fragment is marked with the notification bit in the immediate data field.

On the receiver side, the algorithm supports out-of-order delivery by examining the immediate data of incoming messages to extract sequence numbers and notification flags. Out-of-order packets with sequence numbers beyond the next expected value are temporarily stored in a hash map indexed by sequence number, with boolean values indicating whether each packet carries a notification flag. The algorithm then applies a sliding window protocol that continuously checks for the arrival of the next expected sequence number; when it is received, the algorithm processes it along with any subsequent consecutive packets stored in the hash map, increments the notification counter for those marked with the notification bit, and removes the processed entries from the hash map. This approach guarantees that notifications are triggered only after all preceding messages in the sequence have been received, thereby preserving strict ordering semantics even when packets arrive out of order due to distribution across multiple data QPs.

The algorithm also incorporates a fast path optimization for high-frequency operations, enabling messages to bypass the multi-QP distribution and be sent directly on a dedicated data QP (typically data QP 0). On the receiver side, these messages are processed by directly incrementing the receive-next sequence counter and updating notification counts, eliminating the need for out-of-order tracking. The fast path minimizes the CPU overhead of each RDMA operation, especially useful for algorithms with multiple small to medium RDMA operations such as the inference use case (see \autoref{sec:gpu_resident_collectives}).

%% file: training/zero_copy_eva.tex
\subsection{Evaluation}\label{sec:zero-copy-eva}

This section analyzes the overheads associated with the zero-copy principles. 

We benchmark the point-to-point communication as it serves the building block for various collective algorithms, and compare it with the copy-based approach in baseline NCCL. A zero-copy point-to-point involves a receive buffer address exchange (i.e., handshake), and the data transfer from sender to receiver, following the classical rendezvous protocol. Since the buffer exchange is once per collective, and can be further optimized out if buffer is explicitly reused (e.g., with CUDA Graph in the inference use case, see \autoref{sec:inference}), we exclude it from the reported results.

\begin{figure}[tp]
\centering
\begin{subfigure}[b]{0.45\textwidth}
\centering
\includegraphics[width=\linewidth]{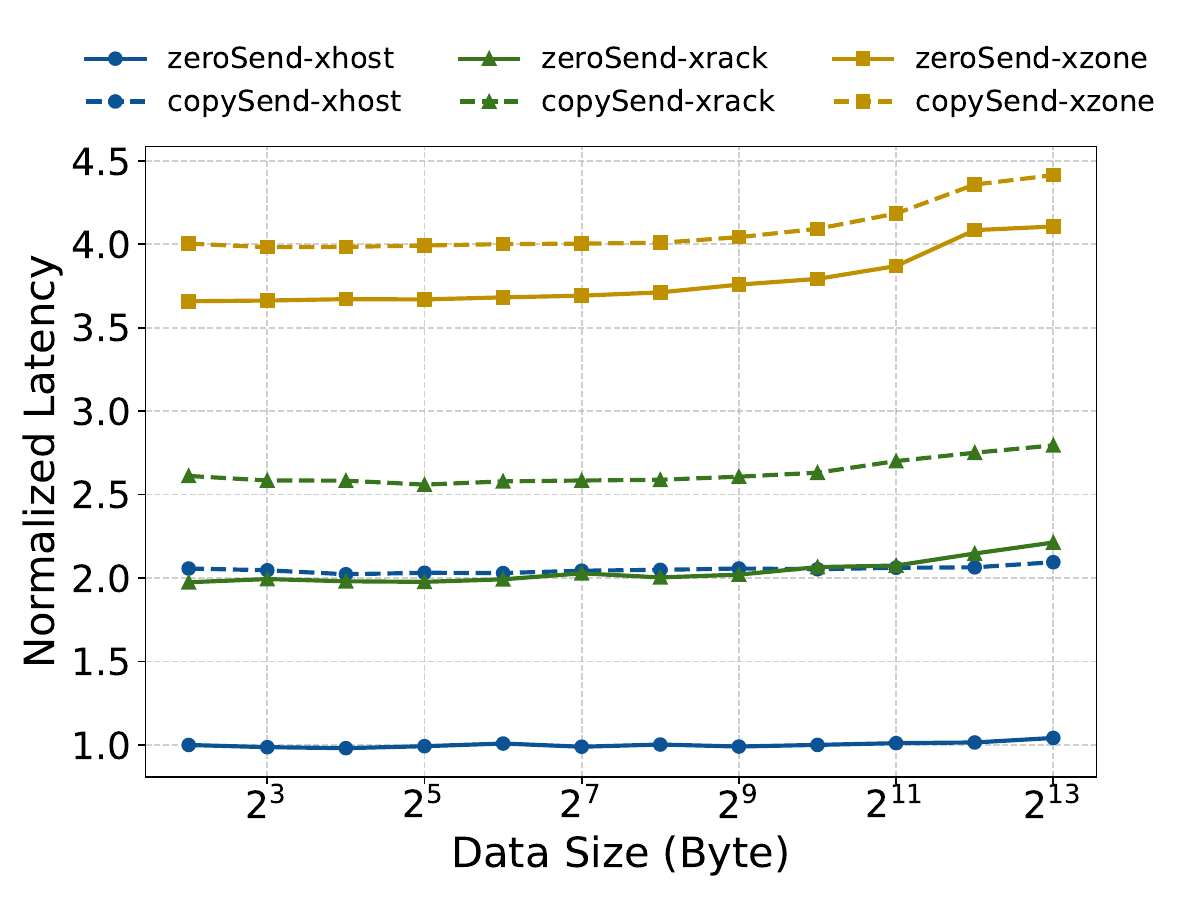}
\caption{Latency under varying network scenarios.}
\label{fig:zero-copy-lat}
\end{subfigure}
\begin{subfigure}[b]{0.45\textwidth}
\centering
\includegraphics[width=\linewidth]{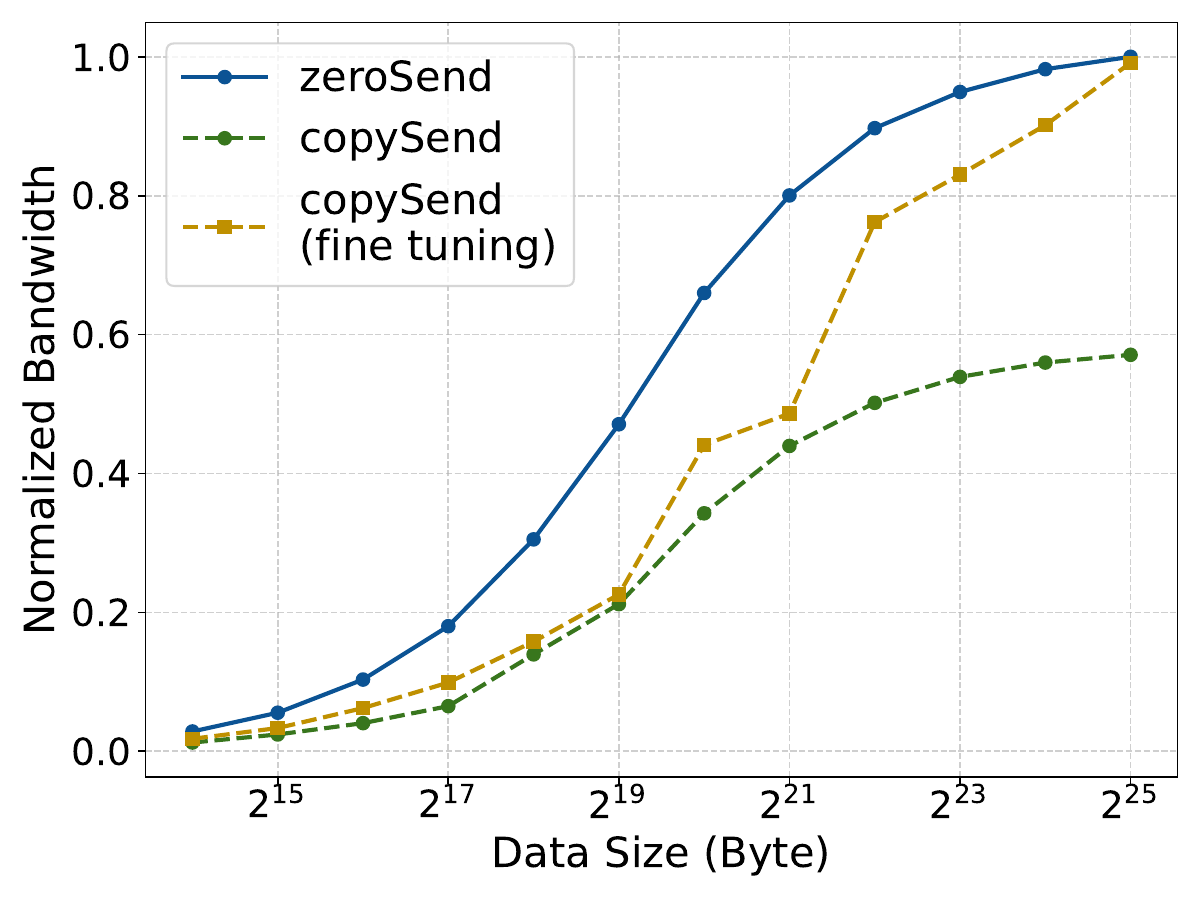}
\caption{Bandwidth with different data size.}
\label{fig:zero-copy-bw}
\end{subfigure}
\caption{Normalized network latency and bandwidth between two GPUs using copy-based and zero-copy data transfer. CopySend leverages baseline NCCL send/receive, and zeroSend is from CTran send/receive without handshake. CopySend (fine tuning) is measured with NCCL\_P2P\_NET\_CHUNKSIZE=1048576, NCCL\_NCHANNELS\_PER\_NET\_PEER=4. In (a), the minimum latency of zeroSend-xhost is normalized to 1. In (b), the maximum bandwidth of zeroSend is normalized to 1.}
\end{figure}

\autoref{fig:zero-copy-lat} and \autoref{fig:zero-copy-bw} compare the zero-copy data transfer and copy-based data transfer between two cross-node GPUs in latency and bandwidth, respectively. For latency, we further detail the results for cross-Host, cross-Rack, and cross-Zone in Meta DC.  Clearly, the copy-based approach introduces a constant tax in latency per transfer due to the additional copy. Such an overhead can increase the overall data transfer time even by 2x in the cross-Host setup. Copy also hurts large message bandwidth due to the chunking at algorithm level. As shown in \autoref{fig:zero-copy-bw}, we cannot achieve reasonable network performance for medium message sizes, even after careful NCCL hyperparameter finetune. In addition, such a finetuned performance can be achieved only at benchmark level but is not practical for production use. This is because some of the hyperparameters have to be applied globally and may degrade other collectives in the same model. For instance, NCCL\_P2P\_NET\_CHUNKSIZE can also affect the performance of MoE AllToAll.

%% file: training/algos.tex
\section{Large-scale Training Customization}
\label{sec:algorithm_performance_optimizations}

Leveraging the flexible zero-copy and host-driven algorithmic framework, we implemented several optimizations and custom features to enhance communication performance in large-scale training scenarios. 

\subsection{PP: Zero-copy and SM-free Send/Receive}
\label{sec:pp-zero-copy-send-recv}

Pipeline parallelism (PP) is a distributed training approach that divides a model’s layers among multiple GPU devices or nodes. This configuration enables simultaneous processing of different microbatches as they traverse each stage of the model. PP makes extensive use of point-to-point send and receive operations, which often span CTSW levels and may cross AI zones. Consequently, achieving low-latency communication over these extended paths requires optimization. Additionally, communication in PP is typically followed by concurrent computation kernels, such as GEMM. In such scenarios, network operations should be tuned to minimize their consumption of GPU resources, thereby reducing contention and enhancing overall throughput.

Initially, we leveraged baseline NCCL's copy-based send and receive operations. We observed two drawbacks: First, the copy-based scheme chunked data into small sizes (128KB-512KB), which was insufficient to mask the high latency of our DC network (i.e., 7x to 15x higher latency at layers of CTSW and cross AI zones compared to in-rack latency, see \autoref{sec:dqplb}).  While increasing the chunk size could mitigate this, it would introduce additional overhead from extra staging copies due to poor pipeline utilization (detailed in \autoref{fig:copy_pipeline}). This would also increase internal GPU buffer usage, thereby reducing the memory available for models. Second, The extra staging copy requires NCCL kernels to occupy GPU streaming multiprocessors (SMs), typically using 4 thread blocks with 640 threads each for a network-only copy-based send/receive. This resource usage slows down concurrent computation kernels such as GEMM.

To optimize network-only communication with computation kernel overlap, completely avoiding GPU SM resources is crucial. We re-implemented the Send/Receive operations following the zero-copy scheme in CTran. In this method, the receiver rank's CPU thread exchanges RDMA registration of the receive buffer with the sender rank. The sender then directly posts an RDMA write from the user send buffer to the remote user receiver buffer. This allows the entire message transfer to be handled by the CPU thread, eliminating GPU resource usage and exposing the full message to the network. Consequently, medium-sized Send/Receive operations (tens of MB) can achieve peak network bandwidth. We note that the user send tensor and receive tensor used in PP Send/Receive operations are often assigned with different underlying memory address range, resulting in high ramp up cost (i.e., overheads to register all touched segments) with tensor auto-registration mode. We applied the memory-pool model for PP to ensure invisible ramp up cost.

\subsection{TP: RMA Put for fine-grained communication and computation overlap}
\label{sec:tp-rma-put}
Unlike network-level bottlenecks, inner-domain collectives present unique challenges in distributed training. Specifically, in Tensor Parallelism (TP), the innermost dimension of parallelism partitions both input and model parameters across devices, which will exchange substantial amount of data during runtime. To achieve high computation efficiency, we leverage TP overlapping to overlap computation and communication effectively, and SM-free data transmission to avoid interference.

Previous studies have explored TP overlapping, such as Transformer Engine from NVIDIA~\citep{transformer-engine}, Pytorch Async TP~\citep{async-tp},  and Flux from Bytedance~\citep{flux}. However, these implementations are restricted to single-host environments (with tensor parallelism degree TP $\le$ 8 on H100 or older platforms) because they depend on CUDA Inter-Process Communication (IPC). Furthermore, solutions such as xFormers~\citep{xformer} and Flux implement overlapping by leveraging device-initiated communications. This approach requires modifications to GEMM kernels, which can potentially reduce computation efficiency, as custom kernels often underperform compared to the highly optimized NVIDIA cuBLAS library.

\begin{figure}[tp]
\begin{subfigure}[b]{0.4\textwidth}
\centering
\includegraphics[height=0.45\textwidth]{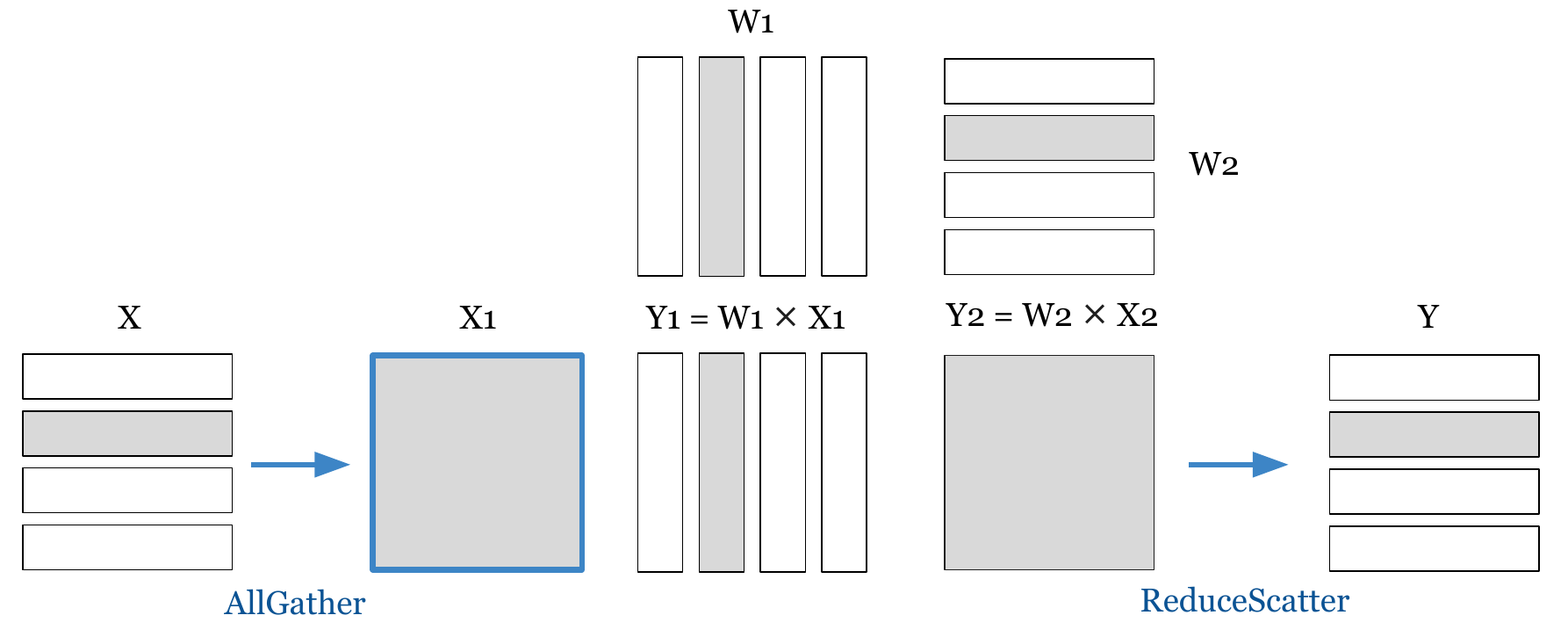}
\caption{Model algorithm overview. In this figure, $X2 = f(Y1)$ where $f$ stands for attention or activation function.}
\label{fig:tp_overview}
\end{subfigure}
\hfill
\begin{subfigure}[b]{0.52\textwidth}
\centering
\includegraphics[height=0.4\textwidth]{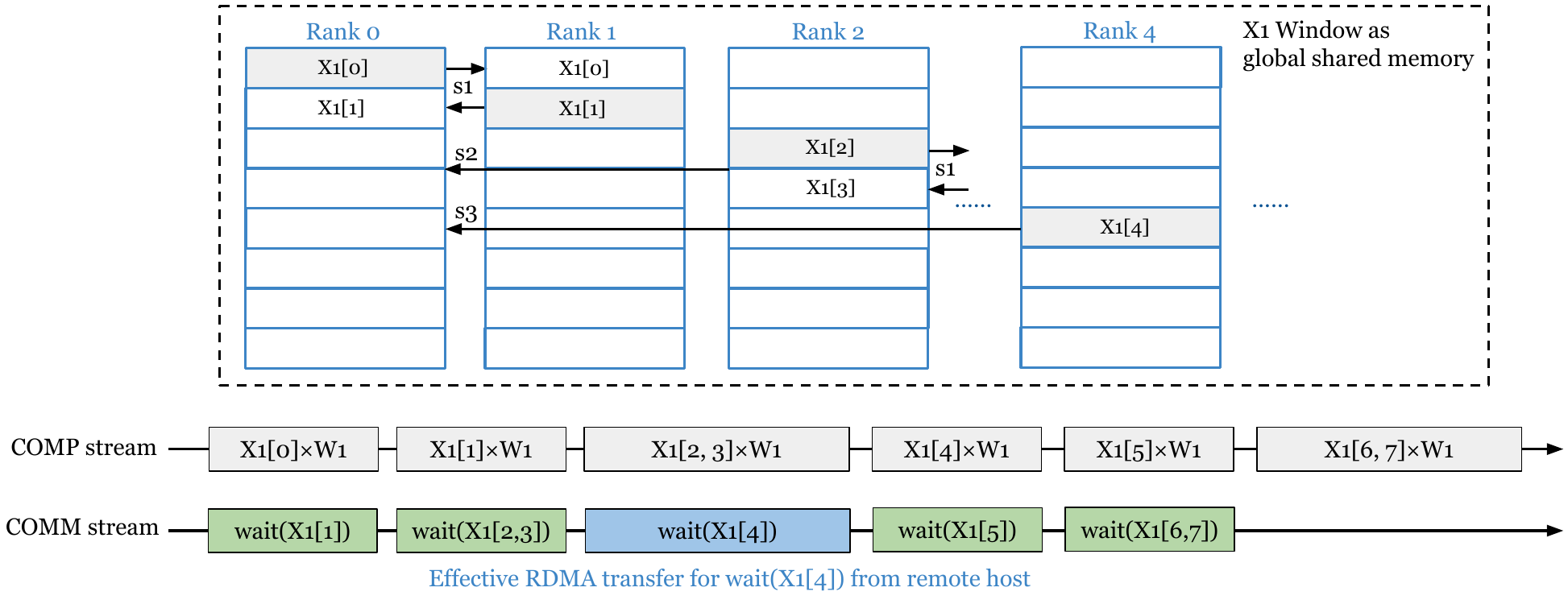}
\caption{AllGather-GEMM Overlap with TP (2 nodes $\times$ 4 ranks). The lower timeline shows the operations on COMP stream and COMM stream of Rank 0. }
\label{fig:tp_ag_pipeline}
\end{subfigure}
\caption{Window and Put-based tensor parallelism compute and communication overlap. X and X1 stands for the input tensor shard on each rank and the AllGather-ed input tensor, respectively. W1 and W2 stand for model parameter shards on each rank in column-wise and row-wise manners, respectively. Y2 is the partial results computed through the two linear layers on each rank, and Y is the final result Reduce-Scatter across ranks. (b) demonstrates the tree AllGather-GEMM pipeline on TP across two nodes with 4 ranks per node. We omit the similar tree GEMM-ReduceScatter pipeline.}
\end{figure}

In Llama training, we utilize the TP paradigm similar to Megatron-LM~\citep{megatron-lm}, which exposes inter-GPU collective operations such as AllGather and ReduceScatter. Our approach to fine-grained TP overlap leverages custom CtranWindow and one-sided Put APIs within CTran, which adheres to MPI-2 standard window and one-sided semantics~\citep{mpi-2}. We provide a detailed description of each design component below.

\textbf{CtranWindow} In CtranWindow, each rank pre-registers a dedicated memory region of identical size and disseminates its corresponding address and access key to all other peers within the same communicator. This design enables any rank to issue one-sided Put operations to an arbitrary peer using the exchanged address and key. The Put API implementation leverages the SM-free CopyEngine for intra-node transfers with NVL and leverages RDMA for inter-node communication. In TP overlapping, we allocate two CtranWindows, to enable overlapping of AllGather and ReduceScatter phases. In the AllGather pipeline, the window buffer receives input tensors from other TP ranks ($X1$ in \autoref{fig:tp_ag_pipeline}), allowing partial GEMM computations to start as soon as the data arrives. A similar pipeline is implemented for the second GEMM operation and the subsequent ReduceScatter phase.

\textbf{Put API.} The Put API abstracts the transfer of data from the sender’s source buffer to the receiver’s destination buffer, essentially mapping to a NVL CopyEngine operation for intra-node transfers or a RDMA write operation for inter-node transfers respectively. The abstraction enhances the programmability of custom model modules by decoupling topology-aware data movement from the higher-level TP overlap logic.

\textbf{Pipeline algorithms.} Thanks to the high programmability of the CtranWindow and Put abstractions, a basic Ring-pipeline overlap can be extended to a \textit{topology-aware Tree-pipeline}, which improves GEMM efficiency by enabling computation on larger tensors in later pipeline stages, while also masking the costly cross-node RDMA transfers through high-speed NVLink chunk transfers. \autoref{fig:tp_ag_pipeline} illustrates the tree-pipeline process from the perspective of rank 0 in a TP configuration with 2 nodes and 4 GPUs per node. Specifically, in the first step (s1), a chunk of size $S$ is exchanged with a neighboring intra-node rank (e.g., rank 0 and rank 1 exchange X1[0] and X1[1]), followed by a GEMM operation on this chunk. In the second step (s2), two chunks are transferred from rank 2, enabling a subsequent GEMM computation on a $2S$-sized tensor. In parallel with the intra-node tree exchange described in s1 and s2, The third step (s3) waits for chunk X1[4] from rank 4 on the remote node. Since the inter-node transfer rate over RDMA is around 8 times slower than intra-node NVL transfer, X1[4] is received after all intra-node communication and computation finished on Rank 0. Therefore, with a setup of 8 H100 GPUs per node, the latency of chunk transfer over RDMA can be effectively hidden by all these intra-node operations.

\subsection{HSDP: Fault tolerant AllReduce}
\label{sec:ftar}

At a scale of 100K devices, training jobs frequently stop and restart due to the high probability of hardware faults. To maintain high training effectiveness—measured as the ratio of productive training time to total runtime (goodput)—elastic training is essential. In data parallel (DP) training, a single failure disrupts the entire DP group because synchronization is required across all workers. Previous work 
enables elastic domain size adaptation when workers have redundant model states. However, our current Fully Sharded Data Parallel (FSDP) scheme partitions both model parameters and optimizer states across workers, which does not have redundant model parameters and weights across workers, and thus making elastic adaptation more challenging.

In collaboration with model researchers, we transitioned the outermost FSDP domain to Hybrid Sharding Data Parallel (HSDP) to better balance fault tolerance and memory efficiency. HSDP adopts a 2D approach: within each inner group (replica group), model parameters and optimizer states are sharded using FSDP, while input tensors are distributed across replica groups. Each replica group independently trains a full model and synchronizes gradients via AllReduce only at the end of each step. This design allows the system to tolerate the loss of gradients from a subset of replica groups during training, improving overall robustness. When a fault occurs within a group, only the affected group shuts down while the remaining groups continue training (shrink phase). Once faulty machines are replaced, a new 4K-GPU group is formed and re-integrated into training (grow phase).

To enable fault-tolerant HSDP, we developed Fault Tolerant AllReduce (FTAR) for robust gradient averaging. FTAR operates alongside a global coordinator, which communicates with replica leads via a dedicated network channel. The coordinator is responsible for fault detection and dynamic group management. When a machine within a replica group fails, the coordinator identifies the affected group and instructs the remaining replicas to continue training with a reduced group size (the ``shrink phase''). Conversely, as new machines become available, the coordinator orchestrates the reintegration of these new machines, expanding the training group accordingly (the ``grow phase'').

The CTran \textbf{host-driven algorithm framework} simplifies fault management, including timeout and error handling, managed from the CPU thread. This framework also facilitates per-step timer and error logging, which is crucial for comprehending performance and faults at scale.

Given that FTARs are designed to communicate across different zones and DC buildings, they often encounter switches with high oversubscription ratios and limited bisection bandwidth, especially when compared to switches within a single zone. To prevent network congestion, it's essential to regulate the number of concurrent data packets transmitted within an FTAR while still maintaining network saturation. The classic Ring algorithm is ideal for this purpose, as it minimizes concurrent network traffic by having each GPU communicate only with its two immediate neighbors in a ring configuration.

\begin{figure}[tp]
    \centering
    \includegraphics[width=\textwidth]{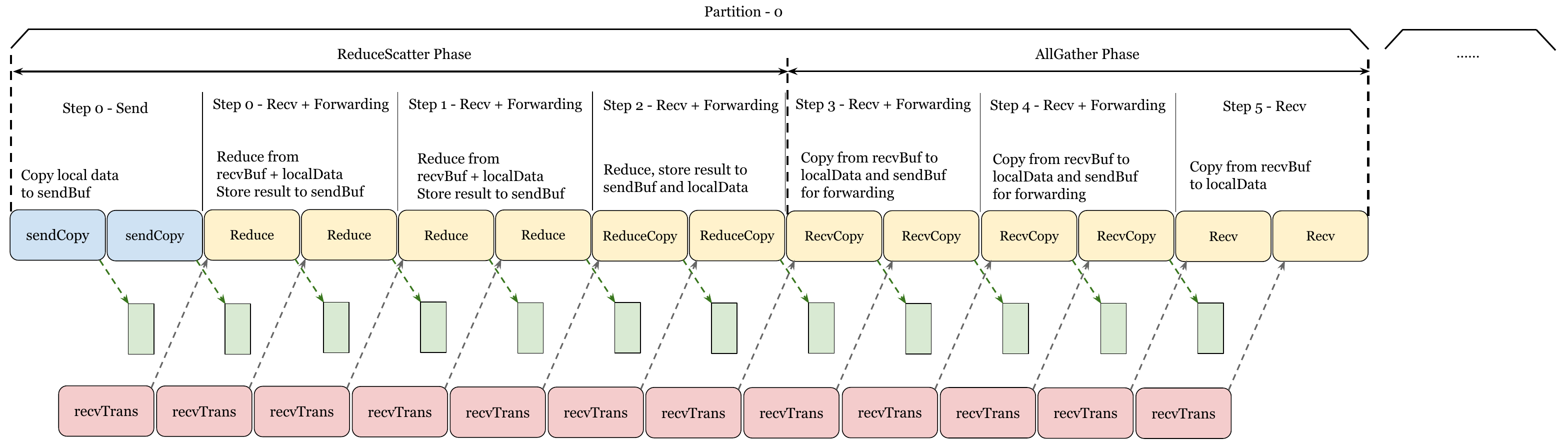}
    \caption{FTAR Ring algorithm with pipelined kernel operations (sendCopy, Reduce, RedCopy, RecvCopy, Recv) and network transfer (recvTrans). The network transfer is tuned to saturate network peak bandwidth and the kernel operations are then assigned with proper number of thread blocks to be fully hidden.}
    \label{fig:ftar}
\end{figure}

FTAR is known to be bound by network bandwidth. Thus, we design a pipeline protocol to hide the in-kernel copy and reduction with network RDMA, as shown in \autoref{fig:ftar}. Unlike NCCL AllReduce, which requires co-tuning of SMs and network chunk size, FTAR uses a fixed chunk size ($S$) and number of chunks ($C$). This approach offers several advantages: First, it provides deterministic concurrent traffic, where a maximum of $S * C$ bytes of concurrent data packets are exchanged between any two peers, ensuring predictability. Second, it separates performance tuning, where the system allows for independent performance tuning of copy/reduce kernel operations and network transfer operations. Developers can optimize the throughput of in-GPU copy/reduce by adjusting the number of thread blocks for a given chunk size, and separately tune network transfer throughput by modifying the number of Queue Pairs and other transport-specific hyperparameters.

With the pipeline foundation established, each kernel step can be finetuned to keep sufficient efficiency at speed faster than the overlapping network RDMA, while minimizing needed GPU SMs. For instance, we combine the reduction and forwarding copy in the ReduceScatter phase of the Ring (named ReduceCopy in \autoref{fig:ftar}), which reduces CPU-kernel synchronization and avoids redundant HBM load. We also avoid unnecessary HBM stores in the intermediate forwarding steps. Furthermore, we enhance instruction level parallelism to maximize the speed of copy and reduction without relying on scaling the number of SMs. 

Putting all the optimizations together, we determined that an 8MB chunk size saturates our network bandwidth and only two thread blocks each with 512 threads are needed to keep the GPU copy and reduction hidden. Further reducing kernel-level overhead offers no benefit if it necessitates more SMs, as AllReduce is already network-bound.

\subsection{Network topology-aware optimizations}

In addition to the model-specific customizations discussed in previous sections, we also implement network topology-aware optimizations for our training jobs, as described below.

{\bf Topology-aware job placement.} Our training job scheduler~\citep{choudhury2024mast} is topology-aware, which assigns consecutive ranks within a training job to nodes that are as close as possible in terms of network distance, thereby enabling the collective communication library to optimize for the actual network structure. Additionally, users can specify constraints for each job regarding the number of GPUs allocated at different network-topology levels (e.g., rack, AI zone, and DC). This flexibility allows users to map specific forms of parallelism—such as tensor parallelism (TP), expert parallelism (EP), and pipeline parallelism (PP)—to designated network-topology levels.

{\bf Topology-aware collectives.} Instead of using non-scalable collective algorithms with linear complexity—such as those based on a ring structure—we adopt latency-hiding algorithms like recursive doubling and halving, which offer logarithmic complexity. For the recursive-doubling implementation of the all-gather collective, we explore several strategies, including nearest-first, farthest-first, and a hybrid approach. Through empirical evaluation, we identify farthest-first as the optimal strategy for our over-subscribed network.

%% file: training/algos_eva.tex
\subsection{Evaluation}\label{sec:algos-eva}

This section reports the benchmark results of selected optimization and custom features for training workloads.

\begin{figure}[tp]
\centering
\begin{subfigure}[b]{0.45\textwidth}
\centering
\includegraphics[width=\linewidth]{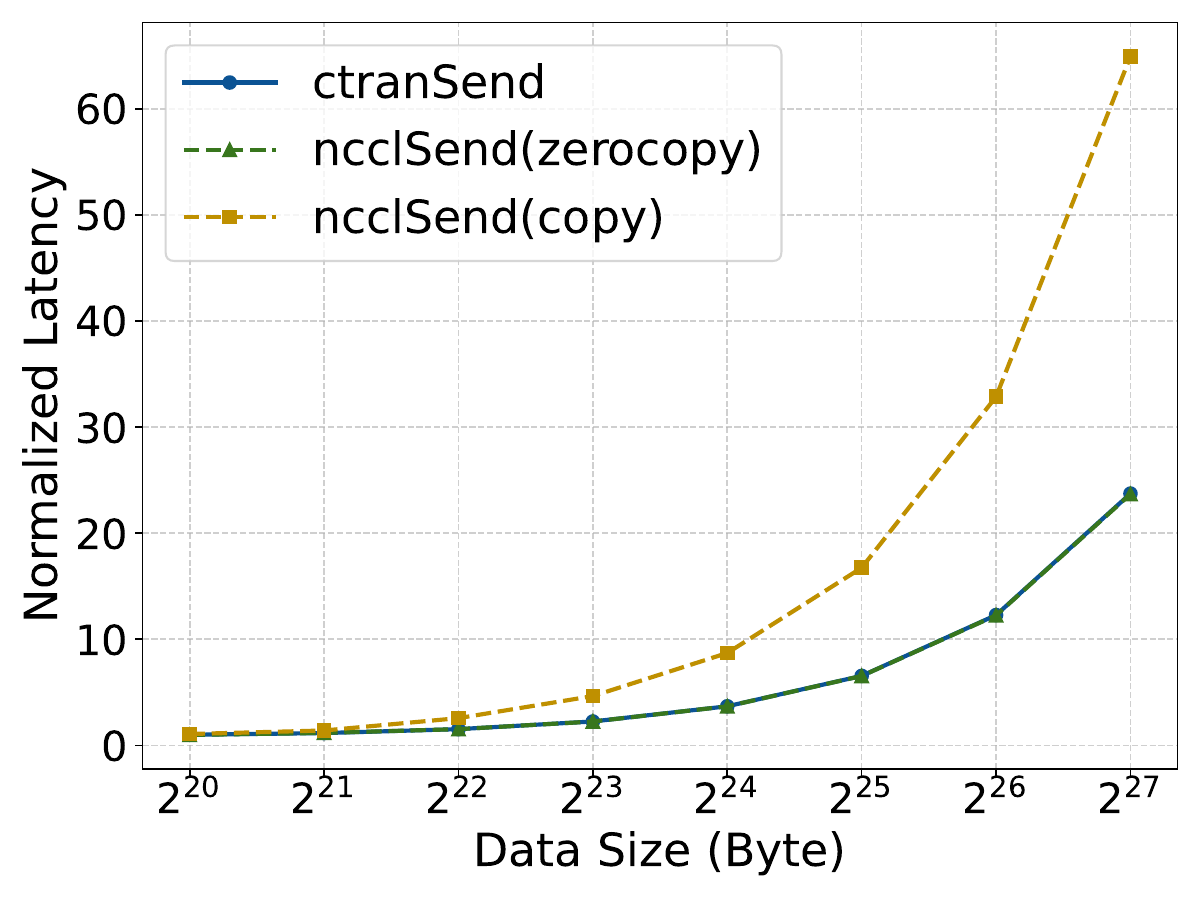}
\caption{Latency.}
\label{fig:p2p-latency}
\end{subfigure}
\begin{subfigure}[b]{0.45\textwidth}
\centering
\includegraphics[width=\linewidth]{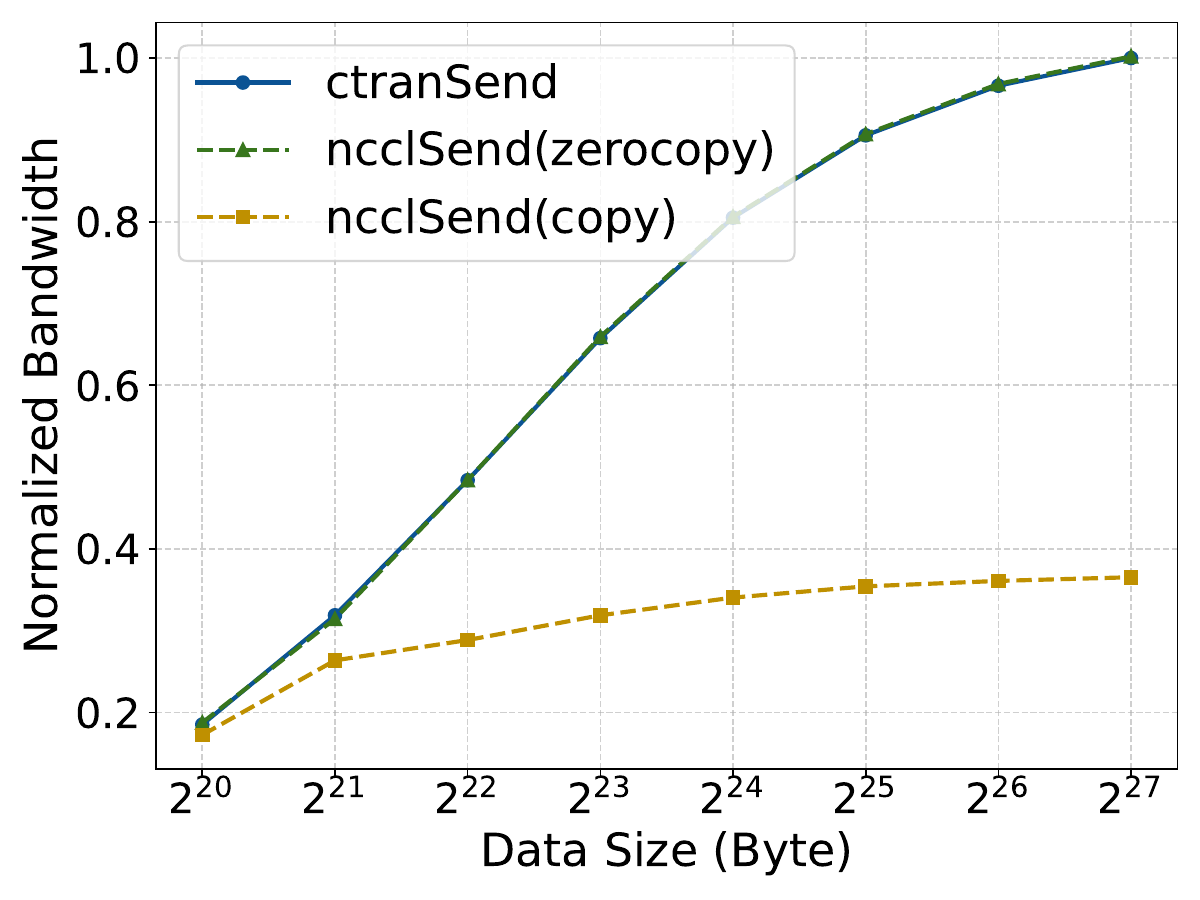}
\caption{Bandwidth.}
\label{fig:p2p-bw}
\end{subfigure}
\caption{Normalized network latency and bandwidth of point-to-point operations with  medium message sizes from 1MB to 128MB used by Pipeline Parallelism. In (a), the minimum latency of ctranSend is normalized to 1. In (b), the maximum bandwidth of ctranSend is normalized to 1.}
\end{figure}

\textbf{Point-to-point.} We evaluate the latency and bandwidth of point-to-point (P2P) operations using CTran zero-copy and compare them against NCCL copy-based and NCCL zero-copy implementations. As shown in \autoref{fig:p2p-latency} and \autoref{fig:p2p-bw}, ctranSend and NCCL zero-copy clearly outperform the copy-based send for the target medium message range, achieving 1.09x to 2.7x speedup. Both of the zero-copy implementations deliver a similar performance, which is expected. We note that, however, we cannot enable NCCL zero-copy in our production workloads due to the suboptimal buffer registration support especially when using with the expandable segment mode of Pytorch cache allocator.

\begin{figure}[ht]
\centering
\includegraphics[width=0.6\textwidth]{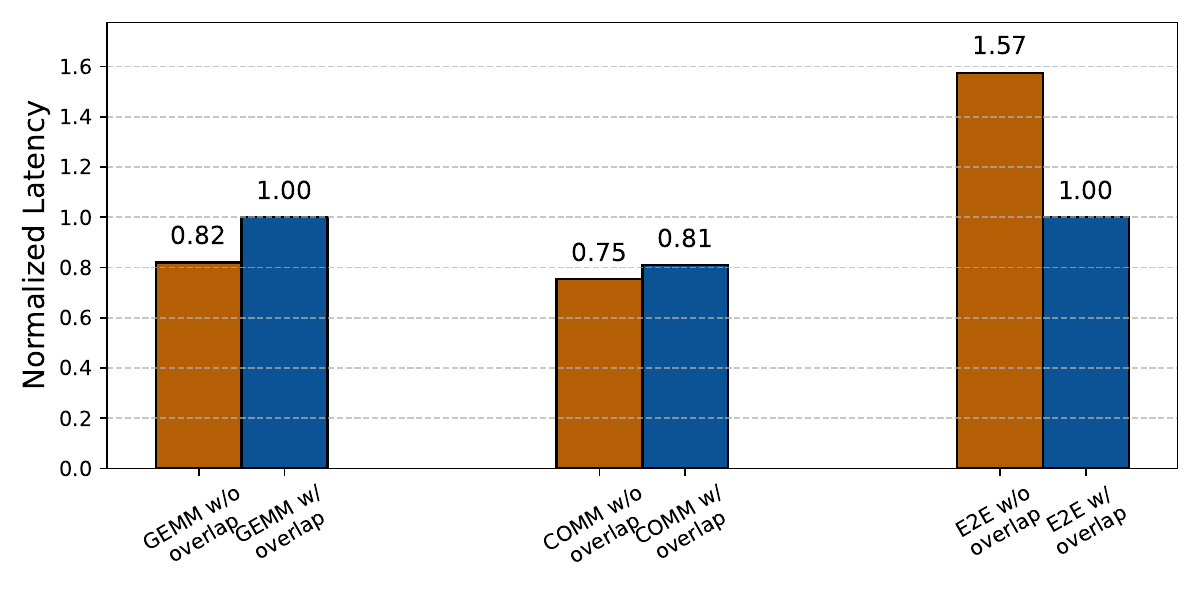}
\caption{The normalized GEMM, COMM and E2E latency of tensor parallelism computation without and with TP-Overlapping on a single node. For both cases, the communication phase transfers 32MB tensors 7 times on each rank. All latency values are normalized with respect to the E2E latency with TP-Overlapping.}
\label{fig:tp-overlapping}
\end{figure}

\textbf{TP Overlapping.} \autoref{fig:tp-overlapping} compares the computation, communication, and end-to-end (E2E) time of a tensor parallelism workload with and without TP-Overlapping on a single node. For the tensor transfer communication time, there is no noticeable difference between TP with overlapping and TP without overlapping. For the computation time, a slight GEMM performance degradation is observed due to reduced computational efficiency with smaller tensor sizes. Since the tensor transfer leverages NVL CopyEngine, it does not consume SM threads and therefore does not interfere with computation. For the E2E time, TP-Overlapping achieves a $1.57\times$ lower latency compared to TP without overlapping, as communication is effectively pipelined within the computation.

\begin{figure}[ht]
\centering
\includegraphics[width=0.85\textwidth]{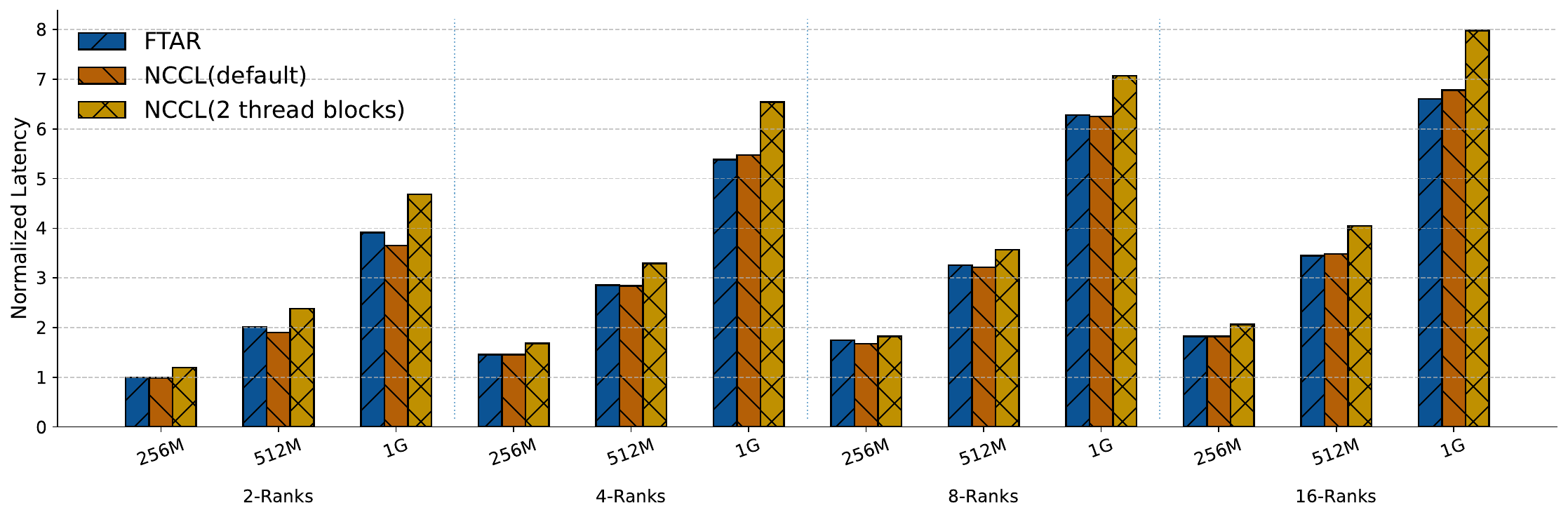}
\caption {Normalized AllReduce latency comparison between FTAR and NCCL with varying rank numbers and message sizes. By default, FTAR uses 2 thread blocks with 512 threads each, while NCCL AllReduce uses 4 thread blocks with 544 threads each. For completeness, we also report NCCL AllReduce results with 2 thread blocks to match FTAR’s configuration. All latency values are normalized with respect to the latency of FTAR with 2 ranks and 256M message size.}
\label{fig:ftar-latency}
\end{figure}

\textbf{Fault Tolerant AllReduce.} We benchmark the FTAR latency across varying message sizes and an increasing number of nodes, and compare it against the standard NCCL AllReduce. As shown in \autoref{fig:ftar-latency}, FTAR achieves comparable latency to NCCL AllReduce, but with only half of the thread blocks (i.e., FTAR uses only 2 thread blocks whereas AllReduce uses 4 thread blocks). If we restrict NCCL to use the same number of thread blocks as FTAR, FTAR achieves 9\%-18\% lower latency than NCCL.
At full training workload, we have also confirmed that FTAR does not introduce any visible interference to the concurrent inner domain computations, thanks to the extremely low SM occupation.

%% file: inference.tex
\section{Multi-node Inference Customization}\label{sec:inference}

Although inference demands less network throughput comparing with training, it requires extremely low-latency to serve user requests. In the meantime, parallelization are needed across multiple GPUs and nodes to enable larger model, faster computation per node and larger batch size. However, achieving both low-latency and parallelization is not easy: the communication between multi-GPUs/multi-nodes brings new challenges to latency, for example, the MoE AllToAll is a well-known expensive collective at LLM inference, even with a few MB of data transfer per operation. It is especially problematic with CUDA graph, which is commonly used in the inference stacks to minimize kernel scheduling overhead. Due to the nature of CUDA graph, extra padded data are transferred, resulting in high latency. 

In this section, we first dive into  the latency issue in the inference workload. We then introduce a GPU-resident communication scheme, which can avoid sending extra data and reduce network transfer volume to the actual needed size, consequently lowering exposed latency. In particular, we dive into the implementation details of AllToAllvDynamic, a first example of GPU-resident collective. Then, we describe a few optimization to improve latency for small message sizes, which is a common use case in inference. Finally, we show end-to-end performance gains with using AllToAllvDynamic.

\subsection{EP: GPU-resident collectives}
\label{sec:gpu_resident_collectives}
Traditional NCCL collective has limitations with ever-evolving ML workloads. Taking NCCL AllToAllv (ncclAllToAllv) as an example. It receives two types of parameters from users: 1) data, which contains the major contents to be sent to peers; 2) metadata, which contains information on how to send and receive data, e.g., send buffer address contains where to read send data, and send counts contains how much data to be sent to each peer. While the data resides on the GPU, metadata resides on the CPU. This can bring new issues: when the collective is enqueued, for example in cuda graph capture mode, metadata is fixed and could not be modified, even if the collective has not started yet. 

\begin{figure}[tp]
\centering
\includegraphics[width=.8\columnwidth]{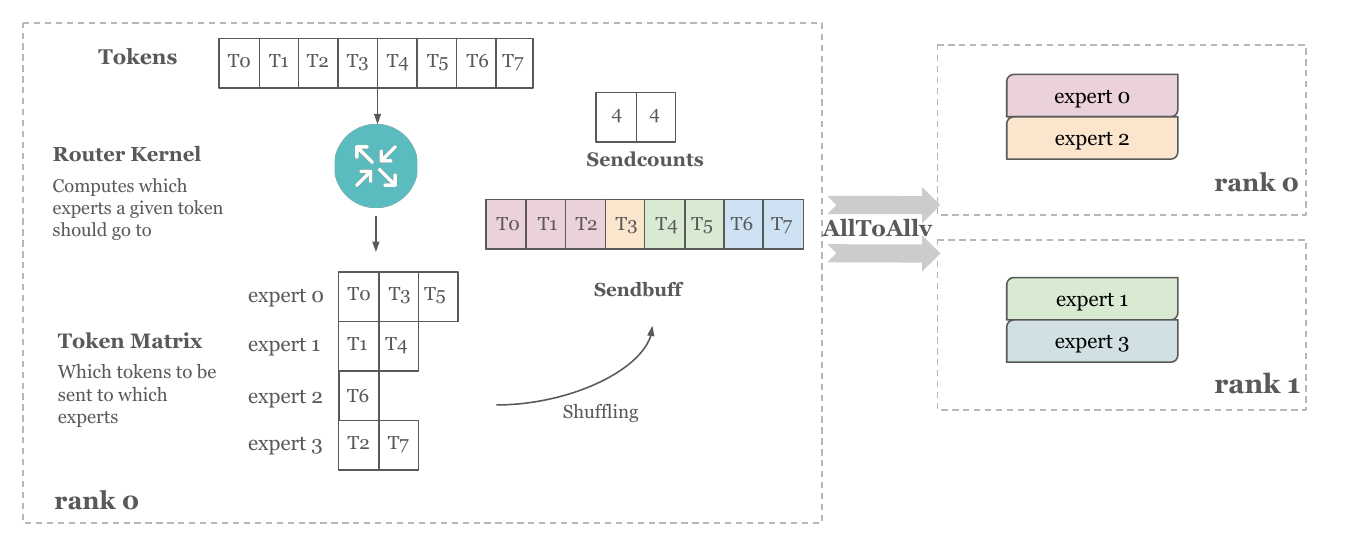}
\caption{Token shuffling workflow. }
\label{fig:a2avd:token_shuffling}
\end{figure}

Let's look into a case study in the inference workload that illustrates how this design can slow down the performance. In particular, we look into the phase of MetaShuffling~\citep{metashuffling} in MoE. MetaShuffling leverage \emph{token choice} mechanism to distribute tokens. As shown in \autoref{fig:a2avd:token_shuffling}, each GPU has a \emph{router kernel} that computes which k experts a given token should go to. The router kernel then generates a result of \emph{token matrix}, which is a map between tokens and experts. After shuffling, a sendbuff (contains the send data) is generated as an input to AllToAllv to send out tokens to peers. 

In this case, the send counts of AllToAllv operations depends on input data and the router kernel’s decision. These send counts cannot be calculated at collective enqueue time (for eager-mode such as in training), or graph creation time (for graph-mode such as inference). This is because at the enqueue time or graph creation time, all the operations (including router kernels and AllToAllv) are enqueued on CPU at the same time, and the router kernel is not started on GPU to finish the calculation. 

\begin{figure*}[tp]
\centering
\includegraphics[width=.8\columnwidth]{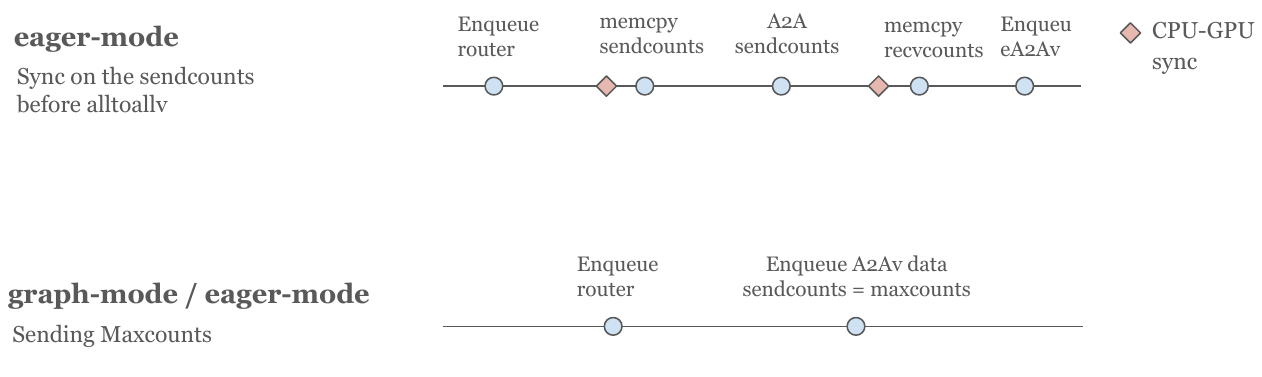}
\caption{Eager-mode verses graph-mode for token shuffling}
\label{fig:a2avd:eager_graph_mode}
\end{figure*}

To resolve the issue that send counts is unknown to AllToAllv, in eager-mode, one can sync on the send counts between GPU and CPU before the start of AlltoAllv as shown in  \autoref{fig:a2avd:eager_graph_mode}. However, the excessive sync between GPU and CPU can bring huge overhead, especially for small kernels. Furthermore, sync is not an option in cudagraph. Instead, we need to send maximum possible counts (maxcounts) as shown in \autoref{fig:a2avd:eager_graph_mode} in cudagraph. The maxcounts need to be large enough to prepare for the worst case where all tokens are assigned to the same expert, which scales linearly with the number of tokens. This means that large paddings with garbage values are sent to peers, and result in longer latency and larger bandwidth usage, greatly affecting performance, especially for inference that requires real-time responses.

\subsubsection{Design Overview}

To tackle these limitations, we introduce GPU-Resident collective, a customized collective whose metadata are resident on GPU. Metadata on GPU allows the input metadata to be modified at any time, up until the collective starts executing. This enables NCCLX to use the actual send counts to transfer data, rather than transferring the maximum send counts, reducing the amount of actual data transferred significantly. 

AllToAllvDynamic is a first example of a GPU-Resident collective focusing on AllToAllv. Other examples (not implemented yet) include GPU-Resident AllGather, GPU-Resident AllGather and AllToAll. In the rest of this section, we take AllToAllvDynamic as an example and overview the design of GPU-Resident collective. 

Let's first look into how traditional NCCL AllToAllv (ncclAllToAllv) works. It cannot receive the message size changes made by the previous kernels because all metadata passed to PyTorch and NCCLX are by-value, i.e., the metadata information is copied at collective enqueue time into internal PyTorch and NCCLX data structures, and hence they cannot be modified afterwards.

\begin{figure*}[ht]
\centering
\includegraphics[width=.75\columnwidth]{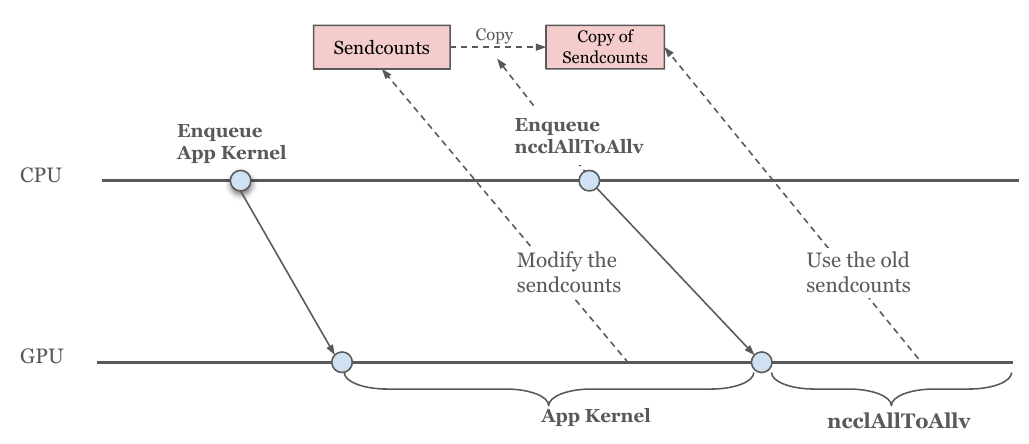}
\caption{Traditional implementation for ncclAlltoAllv. ncclAlltoAllv kernel uses the old metadata (e.g., send counts).}
\label{fig:a2avd:nccl_a2a}
\end{figure*}

As shown in the \autoref{fig:a2avd:nccl_a2a}, ncclAllToAllv makes a copy of the metadata (e.g., send counts) during initlization and enqueue time. After its enqueued on CPU, other applications (e.g., router kernel), which starts before ncclAllToAllv on GPU, can make changes to the metadata on the origional data structure. When ncclAllToAllv starts its kernel on GPU, it can only read the copy of these metdata, and hence can only use the old metadata to perform data transfer. 

\begin{figure*}[ht]
\centering
\includegraphics[width=.8\columnwidth]{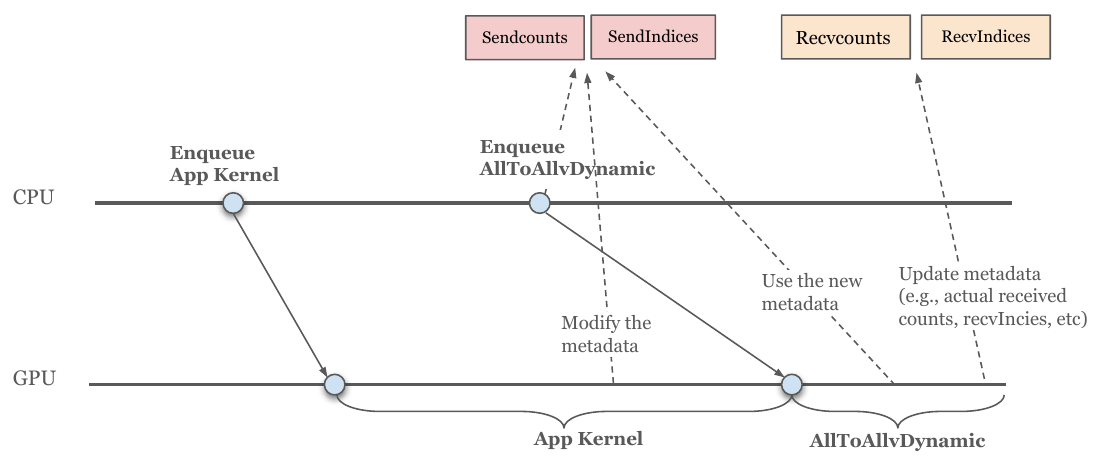}
\caption{AlltoAllvDynamic implementation. AlltoAllvDynamic takes the metadata by reference, allowing its kernel to use the new metadata (e.g., send counts) and updates other metadata (e.g., recv counts) that are needed by the users.}
\label{fig:a2avd:dynamic_overview}
\end{figure*}

In the design of GPU-resident collective as shown in \autoref{fig:a2avd:dynamic_overview}, during enqueue time, instead of copying the metadata, AllToAllvDynamic takes metadata by reference. Even after other application kernel modify the metadata, AllToAllvDynamic can still use the original data structure to read the updated values after it started its execution. Additionally, according to new requirement in the MoE case, we update other metadata such as receive counts and return them back to users. We also introduce a couple of new metadata like send indices, which will be introduced in more details in the next section.

\subsubsection{Implementation Dive-in}
We now dive deep on the workflow and implementation of AllToAllvDynamic. 

\textbf{AllToAllvDynamic workflow:} To exchange data, inter-node CPU side leverages RDMA put developed in the NCCLX framework, and intra-node GPU side launches multi-blocks to copy data in parallel on NVLink. 

\begin{figure*}[tp]
\centering
\includegraphics[width=.8\columnwidth]{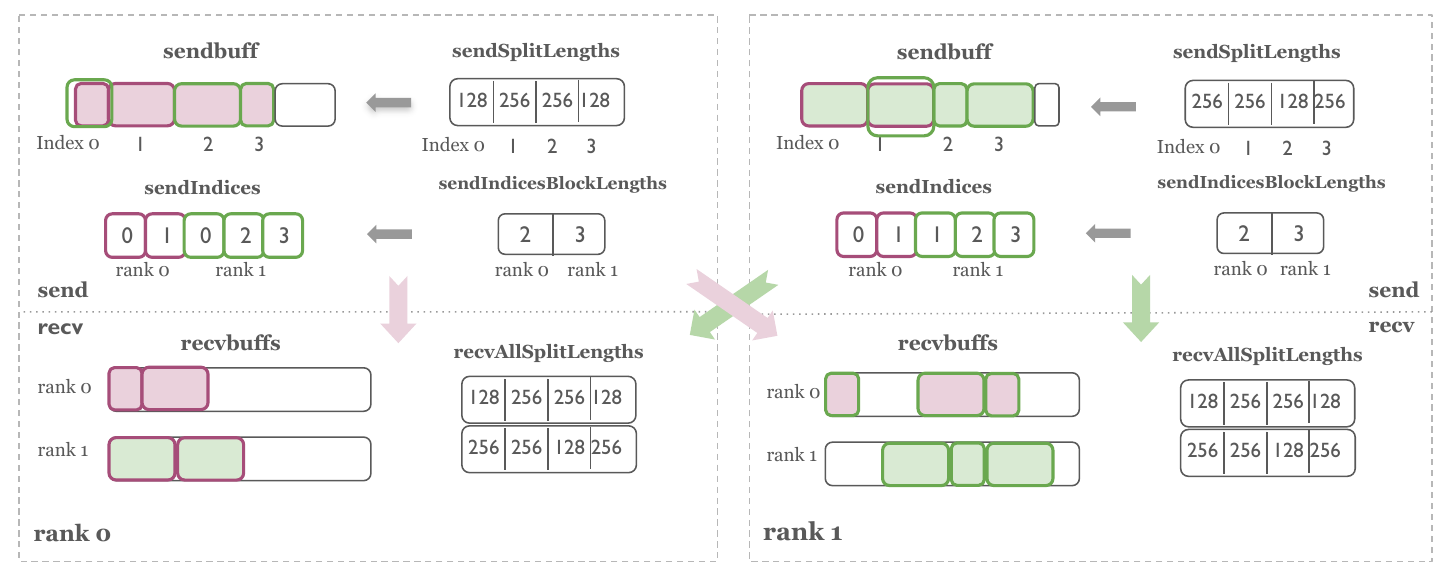}
\caption{AllToAllvDynamic workflow by Example}
\label{fig:a2avd:example}
\end{figure*}

AllToAllvDynamic receives user input metadata and data, which are both resident on GPU. It then split the sendbuffer according to metadata, and exchange both metadata and data with the peer. Figure \ref{fig:a2avd:example} illustrates how AllToAllvDynamic works with an example.  

Each rank receives four parameters for sending the data: \texttt{sendbuff}, \texttt{sendSplitLengths}, \texttt{sendIndices} and \texttt{sendIndicesBlockLens}. \texttt{sendbuff} is a contiguous space on GPU that contains the tokens to send to all ranks. \texttt{sendSplitLengths} $[s_0, s_1, s_2, \dots, s_n]$ denotes how to split the \texttt{sendbuff} into a number of $n$ pieces, where each piece contains $s_i$ number of tokens. \texttt{sendSplitLengths} can be dynamically changed by the previous router kernel and resides on GPU.  \texttt{sendIndices} $[I_0^0, I_1^0, I_2^0, \dots, I_0^1, I_1^1, I_2^1, \dots]$ denotes which indices to be sent to which rank in sendbuff, where $I_j^i$ denotes the split $s_{I_j^i}$ for expert j will be sent to rank i (as expert $j$ is on rank $i$). It is a conceptually 2D array, but flatten into 1D by removing the rank dimension for easy implementation.  The sublist for each rank does not need to contain contiguous indices values. \texttt{sendIndicesBlockLens}  $[l_0, l_1, l_2, \dots]$ denotes how to read \texttt{sendIndices} and resides on GPU, where $l_i$ indicating the number of $l_i$ indices will be sent to rank i, which is equal to the number of experts in that rank (may include the redundant experts). The length of \texttt{sendIndices} is equal to sum of \texttt{sendIndicesBlockLens}.

Taken rank 0 in \autoref{fig:a2avd:example} as an example, the \texttt{sendSplitLengths} splits \texttt{sendbuff} into four pieces, each piece contains $128, 256, 256, 128$ amount of data and labled as indices $\{0, 1, 2, 3\}$ respectively. According to \texttt{sendIndicesBlockLens}, first 2 indices $\{0, 1\}$ in \texttt{sendIndices} are sent to rank 0 and the following 3 indices $\{0, 2, 3\}$ in \texttt{sendIndices} are sent to rank 1. Note that index 0 are sent to both rank 0 and rank 1 due to duplicated experts. 

At the receiver side, each rank gets two parameters that need to be fill in by AllToAllvDynamic and return to the users: \texttt{recvbuffs} and \texttt{recvAllSplitLengths}. \texttt{recvbuffs} contain a list of array that receive data for each rank. In each rank data, the paddings are reserved to be better used as \texttt{sendbuff} in the next round of AllToAllv. For example, on rank 1 receiver side, the data received from rank 0 contains the padding of index 1. \texttt{recvAllSplitLengths} contains the split lengths received from all other ranks. For each rank, it receives all the \texttt{sendSplitLengths} in order to have the padding information. 

\textbf{Implementation challenges and solutions:} There are a couple of implementation challenges we need to tackle. First, when metadata resides on GPU, the CPU cannot access it. However, RDMA operations on CPUs need to read metadata to determine how much data and which pieces of data to be sent to which rank. Second, different from traditional NCCL, the metadata is assumed to be changed up until AllToAllvDynamic starts, which means each rank cannot know how much data it would receive. Such receive counts are used by GPU-side copies and need to be returned to users as well. 

\begin{figure*}[tp]
\centering
\includegraphics[width=.9\columnwidth]{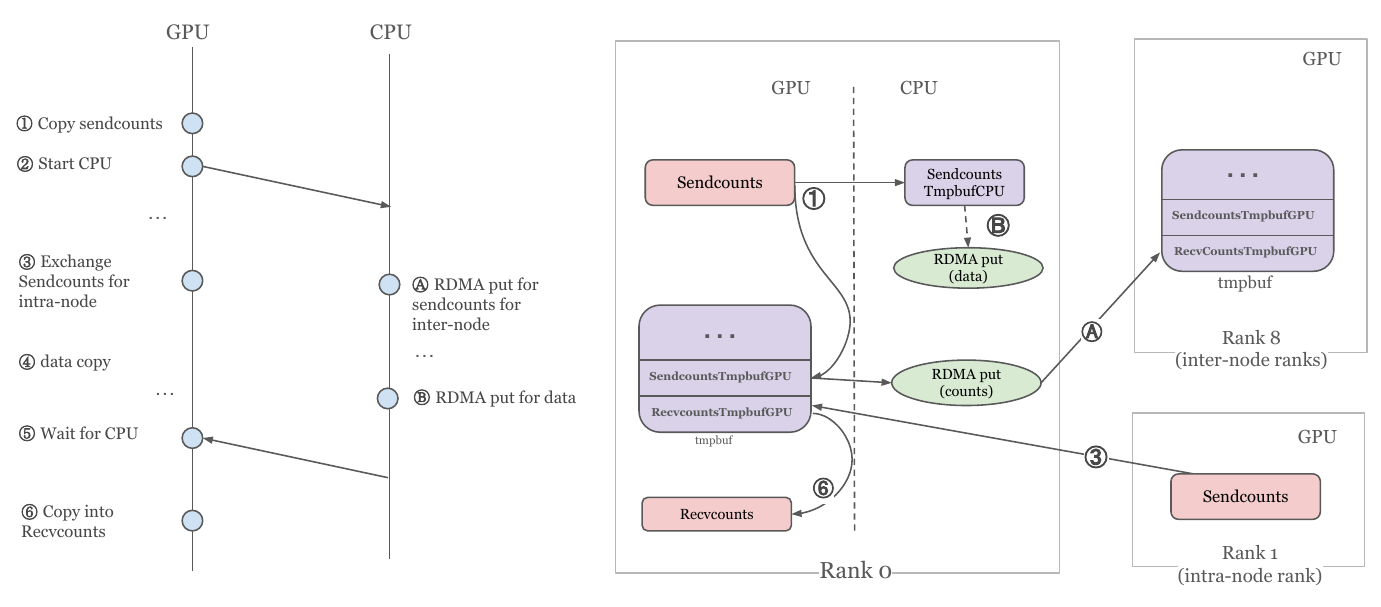}
\caption{Data and Metadata Exchange in AllToAllvDynamic. Tmpbuf are registered temporary buffer used by RDMA. }
\label{fig:a2avd:impl_details}
\end{figure*}

To address the first challenge, we copy metadata from GPU to a CPU buffer. For the second challenge, we exchange metadata (mostly send counts) in addition to data. However, the workflow of these copies is error-prone, and need to be done in a particular order and customized buffers. First, the metadata copy from GPU to CPU buffer need to be done before CPU thread starts, so that the CPU can use the updated values (\textcircled{1}). Second, in addition to copying to CPU buffer, the metadata needs to be copied to a registered GPU buffer (i.e., tmpbuf as shown in Figure~\ref{fig:a2avd:impl_details}) for CPU RDMA put (\textcircled{1}). Third, to receive data, we also need a registered GPU buffer for receive counts (\textcircled{3}). Finally, GPU needs to wait for both CPU and GPU to finish receive the counts, and then copy the received counts back to user buffers (\textcircled{6}). Figure \ref{fig:a2avd:impl_details} illustrate how we exchange metadata in details. 

\subsection{Low-latency Optimizations}\label{sec:low_latency_optimization}

In our workloads, we run RDMA-only AllToAll operations where only one GPU per node participates in the collective. In this scenario, AllToAll communication can become more bottlenecked by CPU overheads than other collective operations. This is because the AllToAll pattern requires issuing RDMA operations to all ranks in the communicator, increasing the CPU's involvement and overhead. For a $N$-ranks AllToAll, the latency can be roughly modeled as $T = T_c * (N - 1) + S / BW$ following the classical LogP model.  We define $T_c$ as the average preparation overhead to issue a message to the remote rank (as known as software overhead), $S$ is the message size per rank, and $BW$ is network bandwidth between two ranks. The implementation in both NCCL and Ctran is to leverage a CPU thread to issue the RDMA, where $T_c$ has to be serialized by $N-1$ times. We assume the RDMA traffic to different ranks are ideally overlapped, thus the overall playload transfer time is $S/BW$. Clearly, when we scale $N$ with small $S$, $T_c * (N-1)$ can become the dominant bottleneck.

Existing work such as DeepEP~\citep{deepep} leverages NVSHMEM to handle such a N-ranks small messages pattern. NVSHMEM is a device-initiated communication model and its implementation can leverage lightweight GPU threads to parallelize the RDMA preparation for different ranks. Hence, the preparation overhead may not increase with $N-1$ and remain flat. Parallelizing via CPU threads, however, may introduce heavier overhead from CPU scheduling and not be the best choice for the host-driven RDMA implementations. Previous work from the MPI community has demonstrated the practical approaches to minimize CPU overhead in small message preparation~\citep{mpi-slow}, and the potential to further parallelize RDMA preparation and issuing via multiple CPU threads~\citep{mtmpi}. Inspired by these study, we focus on reducing $T_c$ in this work. We demonstrate that \textit{the host-driven approach can also achieve low latency similar to the device-initiated model, by carefully optimizing the software implementation}.

\begin{figure}[tp]
\centering
\includegraphics[width=.9\columnwidth]{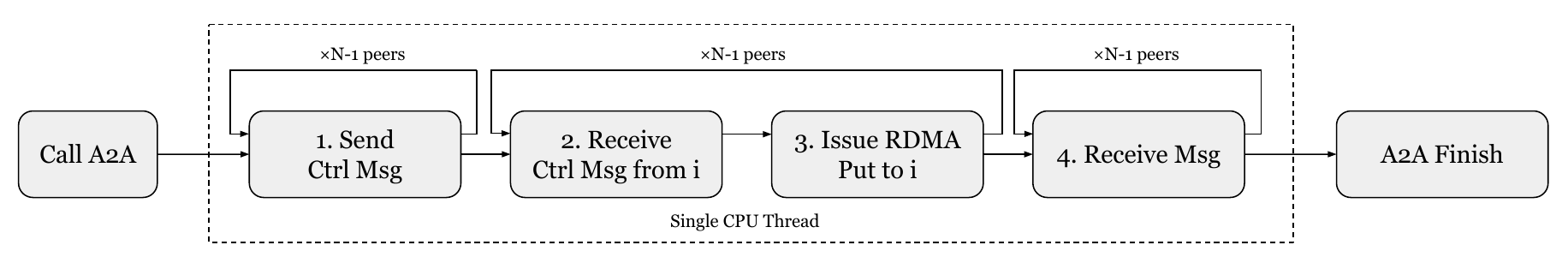}
\caption{CTran AllToAll workflow.}
\label{fig:llc:ctran_a2a_workflow}
\end{figure}

We first breakdown the preparation overhead. We use AllToAll as an example for simplicity; AllToAllvDynamic follows a similar workflow. 
\autoref{fig:llc:ctran_a2a_workflow} shows the zero-copy AllToAll workflow in CTran. First, control messages are exchanged to collect the receive buffer addresses from peers. Then RDMA puts are issued to copy the actual data payload and metadata (if applicable). Finally, each rank wait until it receives the completion notification from all peers. 

\begin{table}[tp]
\centering
\begin{tabular}{|l|c|}
\hline
\textbf{Steps} & \textbf{Latency Percentage} \\
\hline
1-2. Exchange control messages & 50\% \\
3. Issue RDMA puts & 20\% \\
4. Wait for receiving and completing puts & 30\% \\
\hline
\end{tabular}
\caption{CTran AllToAll latency breakdown of $32 \times 8$ (32 H100 nodes, 8 ranks per node) with ($<$ 128KB per rank) message size. Roughly speaking, 1-3 can be quantified as the preparation overhead ($T_c \times N - 1$), while 4 is the actual payload transfer time ($S/BW$).}
\label{tab:ctran_a2a_breakdown}
\end{table}

\autoref{tab:ctran_a2a_breakdown} shows the profiling result for each of these phases for an AllToAll with 8MB message size on 128 H100 GPUs across 32 nodes. To zoom in the RDMA preparation overhead, we explicitly force all messages to be transferred via network RDMA in our profiling study. As expected, the major bottleneck falls into the preparation overhead. Below, we further categorize the main sources of such preparation overhead in our software.~\footnote{Although the preparation overhead is heavily tied with a specific software implementation, the overhead categorization remain generic for most zero-copy host-driven implementations including the zero-copy path in baseline NCCL.}
\begin{itemize}
    \item \textbf{Excessive software complexity on the critical path:} Deep call stacks, unnecessary abstraction layers, fine-grained read/write locks, and redundant checks all contribute to increased CPU time per operation.
    \item \textbf{Control messages exchange:} A typical zero-copy communication starts with a round of control message exchange (often called handshake, shown as steps 1 and 2 in Figure \ref{fig:llc:ctran_a2a_workflow}). The exchange has two purposes: exchanging the memory handle of receive buffers for RDMA transfer, and synchronizing sender and receiver so that sender can ensure the receiver buffer is ready to be updated (e.g., the previous computation consuming the same buffer has finished). This step introduces significant latency, accounting for approximately half of the total AllToAll time in the small-message regime.
    \item \textbf{Inefficient RDMA put operation handling:} The RDMA put path (step 3 in Figure \ref{fig:llc:ctran_a2a_workflow}) incurs high latency due to bookkeeping and load balancing logic that, while beneficial for large messages and high throughput, is unnecessary for small-message transfers. Furthermore, the ibverbs level RDMA post overhead (i.e., calling overhead of the ibv\_post\_send function) becomes visible after optimized out the above mentioned overheads. It is essentially caused by acquiring lock of the ibverbs internal critical section and ringing doorbell to notify the network interface card (NIC).
\end{itemize}

To address the bottlenecks identified in small-message AllToAll operations, we implemented a series of targeted optimizations that span both software design and communication protocol improvements.

First, we focused on reducing software overhead through generic C++ and API optimizations. Functions along the critical path were aggressively inlined to minimize function calling overhead. Additionally, we made the error checks conditional to optionally omit the check in low-latency mode. All low-latency conditions were carefully passed down to the stack via C++ template to avoid any extra branching overhead.

Second, we addressed the significant latency of control message exchanges (steps 1 and 2 in \autoref{fig:llc:ctran_a2a_workflow}) via codesign with the inference workload. Inference workloads often use CUDA graphs to reduce CPU overheads (e.g., CUDA kernel launch). CUDA graph requires the data tensors used in the capture phase remain unchanged during replay~\citep{cuda-guide,pytorch-cudagraph}. Therefore, memory handles can be exchanged once at capture time, and then reused for all subsequent collectives at repeated replay. However, we note that the control message also serves as a synchronization barrier. To remove this barrier, we introduced double buffering to the MoE algorithm so that the two consecutive AllToAlls would always use different receive buffers, avoiding buffer overwriting issue.

Third, we made two optimizations to minimize the RDMA put overhead. We first introduced a small-message fast path to bypass the default bookkeeping and load balancing logic designed for high-throughput scenarios. The fast path directly issues the data as a single RDMA via a dedicated data queue pair. To reduce the ibverbs level post overhead, we  optimized the handling of multiple non-contiguous buffers by implementing work request chaining (as known as scatter list). Chaining allows RDMA transfer from multiple noncontiguous send buffers to be issued together, with the cost to lock and ring doorbell only once. 

We carried the optimizations from AllToAll into the similar AllToAllvDynamic pattern, greatly reduced its preparation overhead and contributed to a higher overall efficiency. Although these optimizations are tackling the challenges in the inference workload, they are generically applicable for all small-message dominated communication. 

\subsection{Evaluation}

\begin{table}[htp]
\centering
\begin{tabular}{|c|c|c|c|c|c|c|}
\hline
 Type & k & \makecell[c]{Batch\\ Size} & host \# & \makecell[c]{Baseline\\ Decode Time (ms)} & \makecell[c]{AllToAllvDynamic\\ Decode Time (ms)} & Improvement (\%)  \\ \hline \hline
\multirow{ 2}{*}{Single Node} & 1 & \multirow{ 2}{*}{128} & \multirow{ 2}{*}{1} & 34.4 & \multirow{ 2}{*}{N/A}  & up to 43.11 \\ 
& 4 &  & & 37.23 &  & up to 0.64 \\ \hline \hline
 \multirow{ 12}{*}{Distributed Inference} & \multirow{ 6}{*}{1} & \multirow{ 3}{*}{128} & 4 & 26.84 & 21.74 & 19 \\ 
 & & & 8 & 29.9 & 19.57 & 34.55 \\  
 & & & 16 & 44.53 & 19.37 & 56.5 \\ 
 & & \multirow{ 3}{*}{256} & 4 & 28.23 & 23.53 & 16.65 \\
 & &  & 8 & 33.77 & 20.23 & 40.09 \\
 & &  & 16 & 50.51 & 20.09 & 59.96 \\
 & \multirow{ 6}{*}{4} & \multirow{ 3}{*}{128} & 4 & 47.57 & 36.99 & 22.24 \\ 
 & & & 8 & 73.82 & 44.09 & 44.27 \\ 
 & & & 16 & 129.85 & 45.45 & 64.97 \\ 
 & & \multirow{ 3}{*}{256} & 4 & 47.5 & 25.71 & 45.87 \\
 & &  & 8 & 85.21 & 25.61 & 69.94 \\
 & &  & 16 & 164.29 & 27.43 & 83.3 \\ \hline
\end{tabular}
\caption{AllToAllvDynamic End-to-End evaluation result.}
\label{tab:a2avd}
\end{table}

We evaluate the end-to-end performance of AllToAllvDynamic, which is equipped with the low-latency optimizations described in \autoref{sec:low_latency_optimization}, and is integrated into the token shuffling stack. We evaluate its improvement on decode time latency. The setup is as follows:

\begin{itemize}
    \item \textbf{Baseline}. Baseline communications contain two Allgather and one AlltoAll, and the compute kernels are the same as AllToAllvDynamic. We enabled CTran in the baseline for a fair comparison. 
    \item \textbf{Comparison dimension}. We focus on testing balanced workload and tune the parameters in the following dimension: token choice 
    $k = \{1, 4\}$, batch size = $\{128, 256\}$ number of host = $\{4, 8, 16\}$. 
    \item \textbf{Methodology}. We omit the result from the first round as it may not be precise during warmup. For each test setup, we run the performance profiling tools for three times and get average numbers across all hosts in all three runs. 
\end{itemize}

\autoref{tab:a2avd} shows the evaluation result. Comparing with single node, there is up to $43\%$ improvement with A2AvDynamic when $k = 1$. Comparing with baseline, there is $15-80\%$ improvement with different setup. With more data transferred (larger k), the gain increases. 

%% file: operation.tex
\section{Other NCCLX Optimizations and Tools}
\label{sec:other}

\begin{figure*}[tp]
\centering
\includegraphics[width=.9\columnwidth]{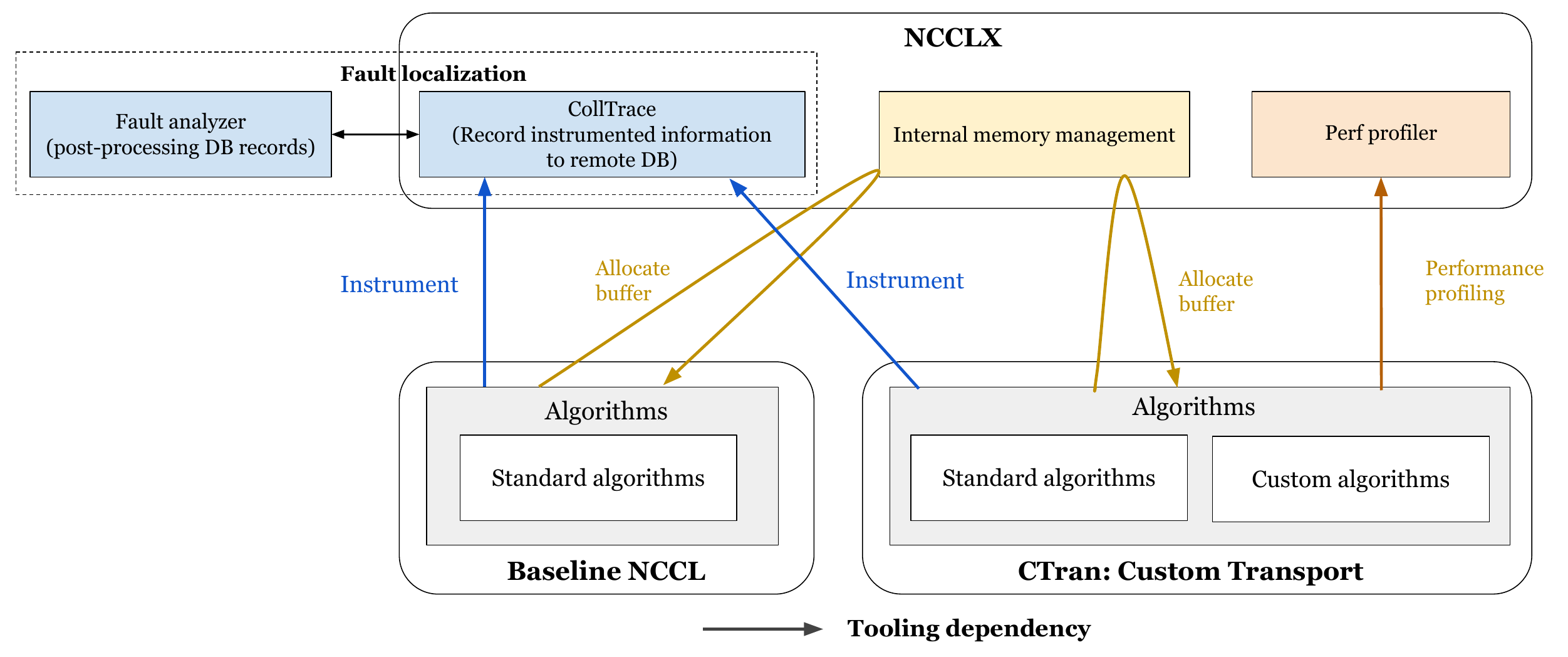}
\caption{Overview of NCCLX tooling within the software stack.}
\label{fig:ncclx_tooling}
\end{figure*}

In this section, we present additional NCCLX optimizations, including scalable initialization for training and enhanced tooling support for resource management, fault localization, performance observability, and CPU emulation. \autoref{fig:scalable_init_control_path} illustrates the control path for scalable initialization in NCCLX, while \autoref{fig:ncclx_tooling} depicts the tooling dependencies across NCCLX, baseline NCCL, and the CTran stack.

\subsection{Scalable Initialization in Training}
\label{sec:scalable_initialization}

Collective communication libraries such as NCCL serve as the backbone of distributed training, facilitating efficient data exchange across thousands of GPUs. At the start of a training job, all GPUs must coordinate to exchange metadata and allocate resources necessary for data shuffling. This coordination is implicitly performed during the creation of torch process groups.

\begin{figure}[tp]
    \centering
    \begin{minipage}[t]{0.48\textwidth}
        \centering
        \includegraphics[width=\linewidth]{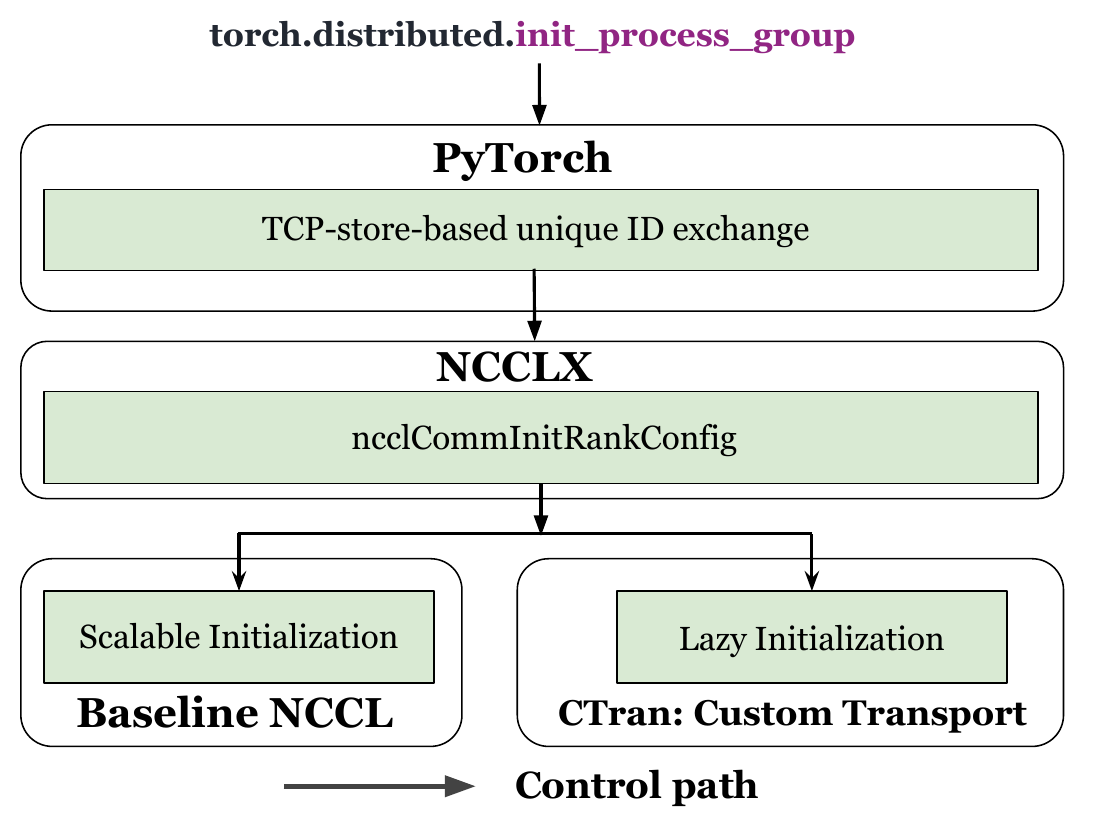}
        \caption{NCCLX scalable initialization control path.}
        \label{fig:scalable_init_control_path}
    \end{minipage}
    \hfill
    \begin{minipage}[t]{0.48\textwidth}
        \centering
        \includegraphics[height=.8\linewidth]{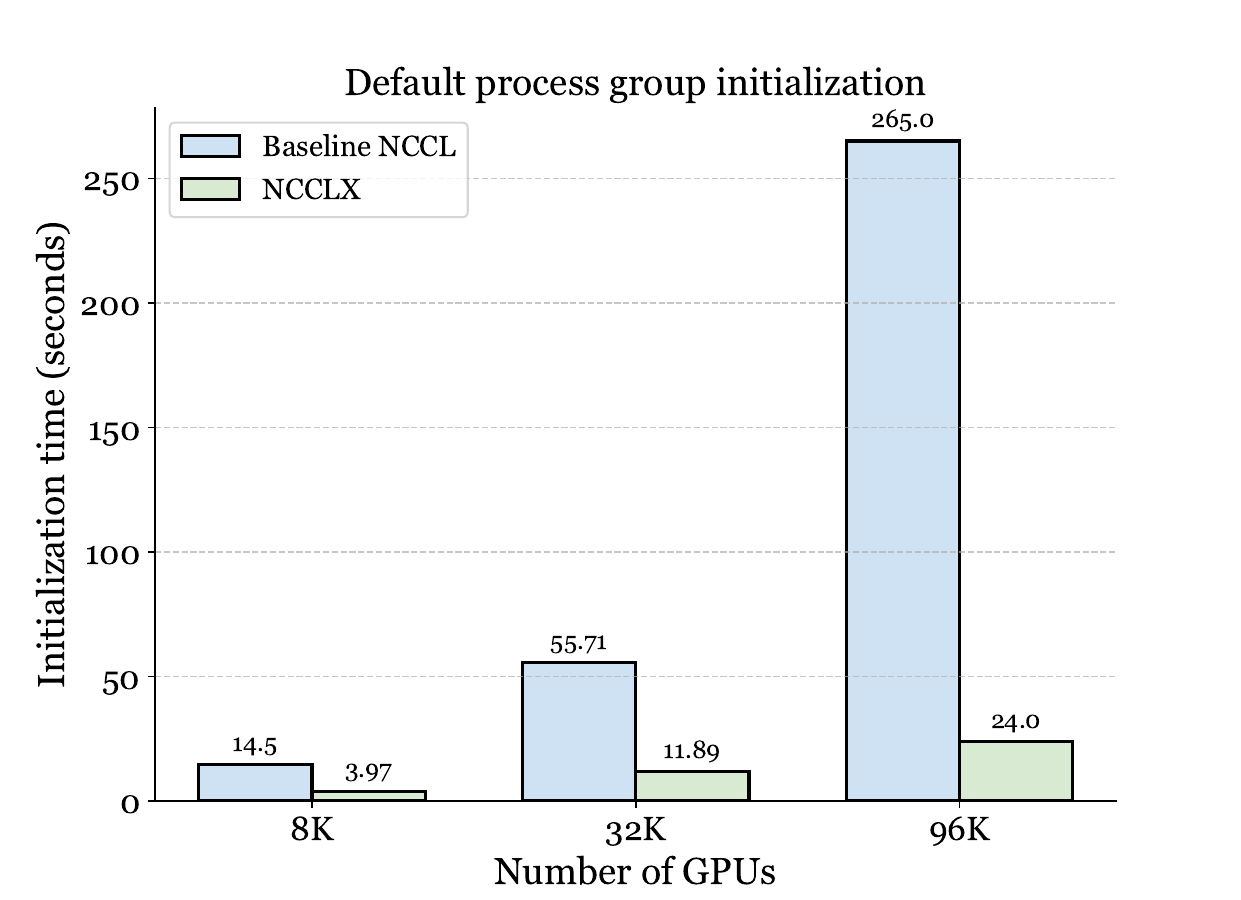}
        \caption{Comparison between baseline NCCL and NCCLX on default process group initialization.}
        \label{fig:bootstrap-tcpstore}
    \end{minipage}
\end{figure}

While training computation scales linearly with resources, communication coordination overhead grows quadratically, making initialization the dominant factor in restart times and overall training effectiveness. At small scales (<1K GPUs), initialization overhead is not concerning, often being in the order of 10s of seconds. However, as we approach 100K and more GPU clusters, initialization time scales non-linearly due to serialized operations, network contention, and computational complexity. Such delays are unacceptable due to high fault rates and frequent job restarts. 

To address this, we carried out series of optimizations across our communication stack including in PyTorch, baseline NCCL and CTran, as shown in \autoref{fig:scalable_init_control_path}. With these optimizations, NCCLX achieves initialization performance improvements of up to 11$\times$ over baseline NCCL.

In baseline NCCL, two phases of the initialization process become increasingly complex at scale before optimization.

\textbf{Bootstrap (\texttt{ncclCommInitRankConfig})} encompasses multiple sub-operations. Each rank initializes local resources (CUDA runtime, topology detection, proxy servers), establishes control channels for peer discovery, and exchanges local state information with all participating ranks. This phase creates the communicator object that enables subsequent data operations. At 100K scale, the last rank waits 100 seconds just to connect to the bootstrap server due to serialized peer discovery operations.

\textbf{Collective Execution (\texttt{ncclAllReduce})} leverages the bootstrap state to execute optimal collective algorithms based on topology, message size, and available bandwidth. The system chooses between ring, tree, or other topological patterns to minimize latency and maximize throughput. 

The initialization technique employed by baseline NCCL presents three challenges:
\begin{itemize}
    \item \textbf{Network-Level Bottlenecks} manifest in multiple ways. TCP connection limitations cause socket queue overflow beyond 64K ranks, resulting in silent connection resets. Bootstrap servers become overwhelmed when handling thousands of concurrent connection requests. The baseline ring formation algorithm creates a serialized bottleneck where each rank must sequentially contact rank 0, causing linear scaling degradation.
    \item \textbf{Computational Complexity Issues} emerge from algorithmic design. Topology computation exhibits $O(N^2)$ complexity, consuming 10s at 48K ranks and projecting to around 100s at 100K scale. Similarly, ring building algorithms scale quadratically, adding significant CPU overhead. These challenges are further exacerbated by process synchronization skewness, where faster-initializing ranks are forced to wait for straggling processes.
    \item \textbf{Resource Allocation Dynamics} shift dramatically at scale. While model loading and data prefetching remain relatively constant, collective initialization time dominates restart overhead. At 96K scale, initialization using baseline NCCL requires over 4 minutes, representing an unacceptable fraction of mean-time-to-failure windows. These challenges necessitate fundamental architectural changes rather than incremental optimizations.
\end{itemize}

NCCLX implements a comprehensive optimization strategy addressing each scalability bottleneck through architectural and algorithmic innovations, described below.

\textbf{Global Process Group}. PyTorch implements two modes of process group (PG) creations, eager and lazy. Traditional lazy mode creates each PG communicator independently, hammering TCPStore with unique ID broadcasts for around 10 PGs per job. NCCLX eager mode creates a single global communicator encompassing all ranks, which is expensive. And then uses \texttt{ncclCommSplit} to derive sub-communicators while reusing global state, thus reducing static overhead per communicator creation. This eliminates repeated bootstrapping and reduces sub-PG creation time significantly.

\textbf{Bootstrap Ring Formation}. Bootstrap topology improvements replace NCCL's centralized bootstrap server with TCPStore-based peer discovery. Meta's TCPStore provides asynchronous I/O optimizations that eliminate the staggered delay of 100-second wait at 100K scale. At 16K scale, this optimization reduces topology formation from 18.45s to 4.1s, with proportional improvements at larger scales.

\textbf{Bootstrap AllGather Optimizations}. AllGather is used in bootstrap phase to exchange control data among participating ranks within a communicator. We optimized AllGather with bi-directional AllGather~\citep{nccl-bidir-allgather} (reducing steps from $N-1$ to $\frac{N}{2}$) and also evaluated tree-based algorithms ($O(log N)$ vs $O(N)$ scaling). In addition, we also reduced number of AllGather calls from 7 to 4 with combined AllGather that enables to shelve off additional overhead during initialization. 

\textbf{CPU Optimizations} transform critical algorithms from $O(N^2)$ to $O(N)$ complexity~\citep{NCCL_PR1789}.
Additional improvements include eliminating unique ID broadcasts 
and optimizing P2P communicator creation for pipeline parallelism.

\textbf{Performance Results}. \autoref{fig:bootstrap-tcpstore} shows the initialization performance results, comparing the baseline NCCL with NCCLX. With the optimizations described above, NCCLX achieves up to 11$\times$ improvement over baseline NCCL performance for the creation of default process group.

In CTran, we employ a different approach to enhance initialization efficiency—a dynamic, on-demand connection strategy. Baseline NCCL is designed for general-purpose scenarios, necessitating the establishment of all connections at startup to accommodate the largest possible collective group. In comparison, CTran’s customization allows us establish connections dynamically and only when needed for specific collectives, reducing unnecessary overhead and improving overall initialization efficiency.

\subsection{Internal Memory Management}\label{sec:resource}

Resource utilization, such as GPU SM and memory and network resources (e.g., Queue Pairs and NIC cache for RDMA communication), is another critical factor during steady training. 
Excessive use of these resources may lead to low training performance, efficiency and, the worst case, failure of training jobs. Specifically, accumulated resources consumption would increase with the number of communicators created by each parallel domain in multi-dimensional LLM training paradigm. 

\autoref{sec:ctran} and \autoref{sec:pp-zero-copy-send-recv} already discuss the SM-free zero-copy transport, this section will focus on other resources. GPU's High-Bandwidth Memory (HBM) is one of the most critical ones for LLM pre-training workloads, which are known to be memory hungry. Typically, the more GPU memory LLM models can utilize, the higher the chance to produce higher performance and quality results because it enables larger batch sizes and more hyperparameter spaces to explore and tune. However, it is fairly common that LLM pre-training jobs encounter out-of-memory (OOM) errors due to the limited available GPU HBM, e.g., Nvidia GPUs H100, which provides around 60~GB available HBM out of 80~GB.

Communication libraries such as NCCL are often optimized for absolute performance of collective communication and overlook resource efficiency. NCCL, the de-facto communication library on Nvidia GPUs, allocates a significant amount of internal GPU buffers, which are not shared with applications, to provide the best possible pipeline designs for various hardware topologies and collective patterns. However, this leads to a significant waste of memory in today’s multi-dimensional parallelism of LLM training. In early experiments with Llama4 pre-training, we have seen NCCL consume approximately 10~GB HBM (which is about 12.5\% of H100 HBM) across 10+ parallelism groups, each associated with a NCCL communicator that allocates dedicated memory. 

There are three fundamental design choices causing resource inefficiency in NCCL: 
\begin{itemize}
    \item \textbf{Eager resource allocation:} To ensure high-performance pipeline for the copy-based communication, NCCL allocates internal buffers and QPs per send/recv peer and per NCCL protocol (LL/LL128/SIMPLE), per communicator. It can easily adds up when more number of communicators are created, especially for GPUs performs All-to-all communication within NVLink domain. Additionally, different algorithms (RING vs TREE) allocate dedicated resources at runtime, but not always used during the training.
    \item \textbf{Multi-channel Designs:} NCCL implements a concept of multi-channel to improve the concurrency of data movement and HW/network utilization. Each channel allocates dedicated resources, as mentioned above, to ensure the best performance. However, it leads the waste of resources when a communicator does not require a high number of channels to maintain high performance, e.g., only perform small-message collectives.
    \item \textbf{Store Metadata on HBM:} NCCL ``channel'' contains small metadata (FIFO status, peer connection info, etc.) and will consume 2~MiB per channel due to the alignment requirement of CUDA low-level virtual memory management interfaces (e.g., cuMem APIs). Although it achieves low latency when NCCL kernel loads these metadata, it adds extra pressure on HBM and memory fragmentation/waste. More importantly, such metadata also scales with the number of ranks in a communicator, which can accumulate to more than 1~GB at 100k GPU scale.
\end{itemize}

To tackle these shortcomings for LLM trainings, NCCLX optimizes the resources efficiency while keeping high-performance collectives by implementing following new features:
\begin{itemize}
    \item \textbf{Lazy Algorithm Initialization/Connection:} Only allocation resources and connect peers when an algorithm is used at runtime. Similar idea apples to CTran introduced in~\autoref{sec:ctran} as well.
    \item \textbf{Lazy Channel Allocation:} Only allocate minimal required channels and grow at runtime as needed, with the factor that some communicators only perform small collectives in the lifetime of a training job, and do not require multi-channel concurrency to saturate network bandwidth.
    \item \textbf{Slab Allocator for Metadata:} Implement a slab allocator to store metadata, multiple channels' metadata can fit into a same GPU page to reduce the memory waste and fragmentation. 
\end{itemize}

After these optimizations, we were able to reduce NCCL GPU memory usage by almost 2x among 10+ communicators in large scale as \autoref{tab:memory_savings} shown as well as production runs up to 100k GPU. It is worth mentioning the new features also reduce number of QPs within 2000 per NIC, that typical Mellanox CX-7 or newer HCA can handle without performance regression.
Next, we discuss more details of these features.

\begin{table}[tp]
    \centering
    \renewcommand{\arraystretch}{1.2}
    \resizebox{\textwidth}{!}{
    \begin{tabular}{|l|l|l|}
        \hline
        \textbf{Feature} & \textbf{CVAR to Enable} & \textbf{Saved HBM per GPU for Llama4 Pre-training} \\
        \hline
        Lazy algorithm connect         & \texttt{NCCL\_LAZY\_CONNECT=1} & 8.09\,GB $\rightarrow$ 6.65\,GB \\
        Ctran lazy connect                     & \texttt{NCCL\_LAZY\_CONNECT=1} & 6.65\,GB $\rightarrow$ 5.86\,GB \\
        Slab allocator                         & \texttt{NCCL\_MEM\_USE\_SLAB\_ALLOCATOR=1} & 5.09\,GB $\rightarrow$ 4.7\,GB \\
        Lazy channel allocation                & \texttt{NCCL\_LAZY\_SETUP\_CHANNEL=1} & 5.39\,GB $\rightarrow$ 4.2\,GB \\
        \hline
    \end{tabular}
    }
    \caption{Example of memory saving by the proposed features for Llama4 pre-training on 64K GPUs}
    \label{tab:memory_savings}
\end{table}

\textbf{Lazy Algorithm and Channel Allocation} 
NCCL allocates the maximum amount of memory as well as peer connections statically at communicator initialization time, mostly based on topology detection. For example, NCCL would allocate more memory/channel for communicators with peers within NVLink domain, since it can utilize more buffers and concurrency to saturate NVLink's high bandwidth. 
However, as NCCL's eager allocation attempts to allocate the maximum resources it requires at initialization time since it has no knowledge of the communication patterns at runtime, leading to significant waste of resource. 
Especially worst for memory, as each parallelism group is designed to perform particular collectives within specific message sizes based on the model configurations and scales. For example, DP group performs Allgather and ReduceScatter, which only use Ring algorithms and no need to allocate resources for Tree algorithm.

First, NCCLX enables a lazy connect feature that dynamically allocate resources and connect for each algorithm only when its first use at runtime~\footnote{We shared our optimization insights with NVIDIA NCCL team that co-contributes a similar feature in the NCCL library since v2.22 release.}. In our empirical study of llama4-like pre-training runs, this feature alone saves about 2.8~GB among 10+ NCCL communicators as demonstrated in~\autoref{tab:memory_savings}.

\begin{figure}[tp]
\centering
\includegraphics[width=.6\columnwidth]{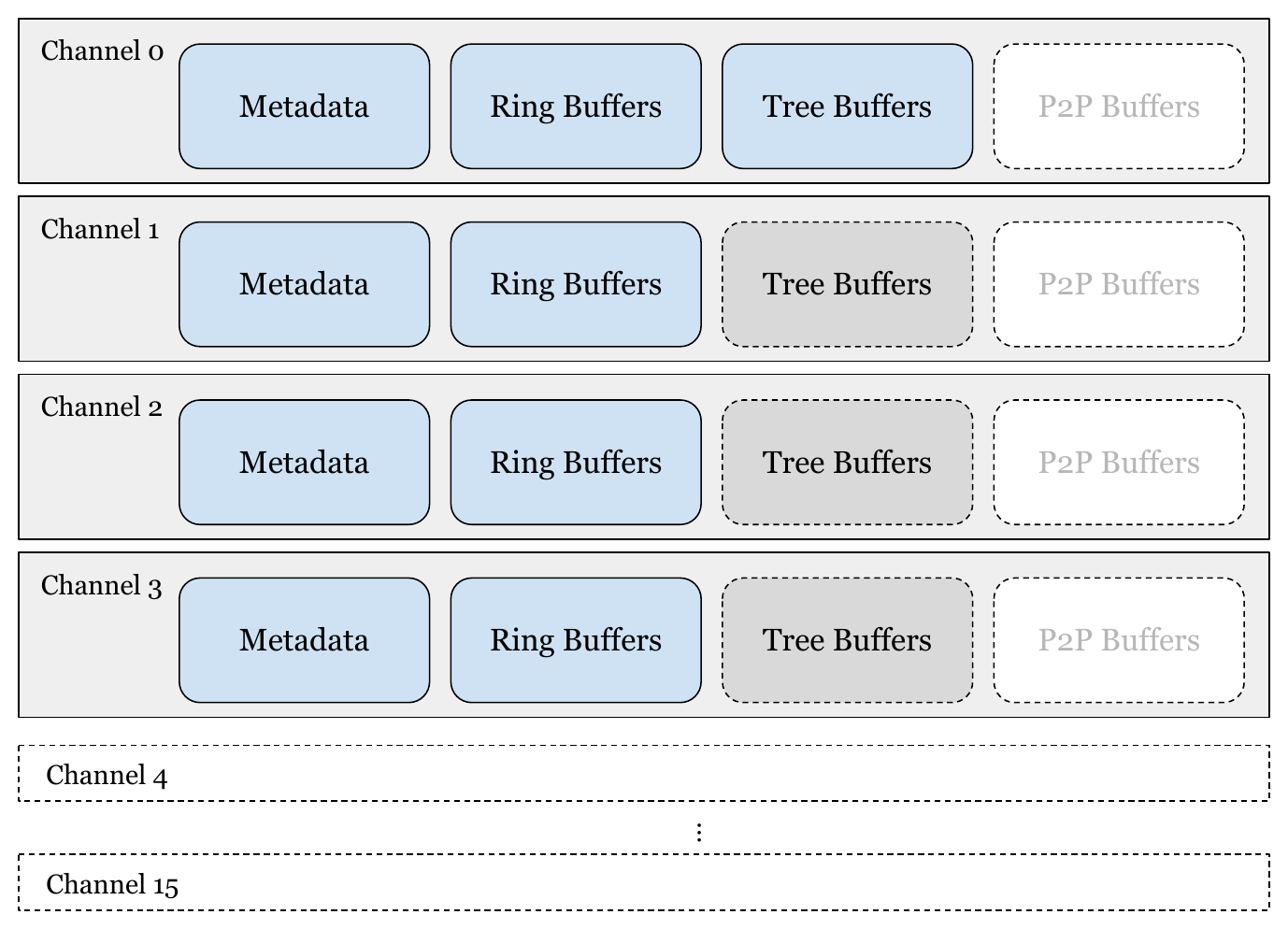}
\caption{Lazy Feature Example; memory can be saved by allocating fewer channels as well as fewer metadata/algorithm buffers}
\label{fig:lazyChannel:example}
\end{figure}

Next, NCCLX further introduces a lazy setup channel feature to lazily allocates channel resources. In addition to lazy connect feature, which only allocates resources for specific algorithm on all channels, this feature avoids allocating unnecessary channels at all. The motivation of this feature is based on the observation that some communicators only perform small collectives during the LLM training and do not require multi-channel concurrency. As the example illustrated in \autoref{fig:lazyChannel:example}, if a collective only requires 4 channels to perform Ring-based algorithm, NCCLX will not allocate resources for the rest of 12 channels at all (assuming it uses 16 channels by default). Similarly, if a communicator only performs Tree algorithm for small Allreduce at runtime, NCCLX can only allocate Tree buffers on 1 channel compared to baseline NCCL would allocate resources in all channels and only use it partially. Note that NCCL runtime already implements tuning logic to decide the number of channels and what algorithm when enqueuing a collective.
As presented in as \autoref{tab:memory_savings}, this feature saves around additional 1.2~GB of memory in our empirical study of llama4-like pre-training runs.

\begin{figure}[tp]
\centering
\includegraphics[width=\textwidth]{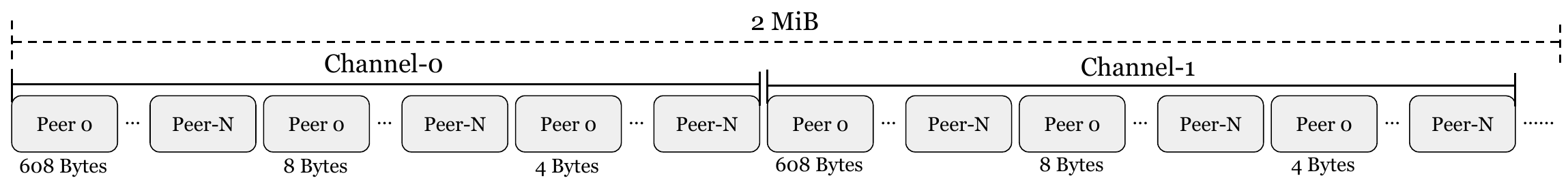}
\caption{Example usage of Slab Allocator; multiple metadata can fit into 2MB slabs to avoid waste}
\label{fig:slabAllocator:example}
\end{figure}

\textbf{Slab Allocator for Metadata}
Finally, to reduce GPU HBM usage at large scale only for NCCL's Metadata (roughly 600 Bytes per peer per communicator based off NCCL v2.27, as of this manuscript is written), we implemented a \texttt{Slab Allocator} to concatenate metadata from multiple communicators into contiguous GPU pages as depicted in~\autoref{fig:lazyChannel:example}. For example, a 2~MiB single GPU page can hold metadata for 3000 peers. 
All slots are freed at communicator destruction time in the end of the training job. This feature saves about 400~MB of memory in our empirical study of llama4-like pre-training jobs as \autoref{tab:memory_savings} shown.

\subsection{Fault localization}
\label{sec:fault_analyzer}

In distributed training, hardware failures—such as faulty network interface cards (NICs) or unhealthy GPUs—often manifest as job failures or stuckness. Due to the synchronous nature of model training, a single defective hardware can cause the entire job to fail. Consequently, the mean time between failures (MTBF) for training jobs decreases as the scale of training increases. At scales involving several thousand GPUs, interruptions occur with such frequency that manual troubleshooting becomes impractical. Production engineers may spend hours or even days identifying and isolating the problematic machine, during which additional failures may arise. To mitigate this operational challenge, we developed Fault Analyzer, an automated system that analyzes training job failures and efficiently localizes faulty machines within distributed training jobs.

Since version 2.24, NCCL has introduced the reliability, availability, and serviceability (RAS) subsystem to facilitate the diagnosis and debugging of crashes and hangs. The RAS subsystem initiates a dedicated thread for each NCCL process, enabling the exchange of information among RAS threads and monitoring the health of each process. When a job hangs or fails, RAS assists troubleshooting by providing status updates for each communicator and diagnostic information for potential failures, such as mismatches in collective operation counts.

However, based on our experience with large-scale training, we have identified two key limitations of the RAS subsystem: {\em (1) Limited Coverage of Cascaded Failures:} The complex, multi-dimensional model parallelism design in large-scale training often results in failures within a collective operation in one dimension propagating to communicators in other dimensions. This typically manifests as certain ranks failing to schedule the latest collective or collective kernels waiting for prior operations in the stream to complete. RAS does not adequately distinguish between original and cascaded failures in these scenarios. {\em (2) Insufficient Kernel Tracking for Diagnosing Hangs:} Empirically, tracking whether a collective kernel has started is highly valuable for diagnosing hangs in large-scale training, as it helps identify the first collective operation that encountered a failure. However, RAS currently lacks the capability to monitor the start and end of collective kernels.

We instrumented the NCCLX library with tracing instructions at both per-collective and network RDMA granularity to capture detailed state information (CollTrace). This tracing forms the foundation for subsequent analysis by providing comprehensive insights into the behavior of NCCLX during distributed training.

We developed collective interdependency-aware analysis rules that efficiently detect the initial stalled collective operation and localize the faulty host within LLM training jobs involving multi-dimensional model parallelism. This approach enables precise identification of failures in complex distributed environments.

Due to the lack of full visibility of GPU operations outside of NCCL, we built our a Fault analyzer that detects the inter-collective dependencies based on two assumptions: (1) When analyzing the job, the job has hung for a sufficiently long time to finish all collectives that could be finished. (2) If a collective kernel did not start on a given rank, it is (directly or indirectly) waiting for the running collective on that rank.

Based on those two assumptions, we built the analysis logic to detect dependencies between collectives as follows: On a given global rank G1, we may have multiple collectives A, B, C. If is A is active, and B and C are pending, then we assume B and C are blocked on A. By checking the collectives that are not waiting on anything else, we can find out the collective first encountered failure.

To demonstrate the practical utility of our Fault Analyzer, we present two representative scenarios encountered during large-scale model training: hardware failures and model code issues. These examples illustrate how the system streamlines triage and resolution.

\textbf{Hardware Failure Triage} During a training job spanning 8K GPUs, the system experienced repeated job restarts. The error logs provided by the training framework were insufficient to immediately identify the root cause, complicating the troubleshooting process.

The Fault Analyzer streamlined the triage process: First, it automatically aggregates collective and network operation traces. By analyzing inter-collective dependencies using collective tracing, we found that the last All-Reduce in the second Data Parallelism Process Group was the first collective to fail.
Second, it Analyzes network operation traces. By examining the network operations in NCCLX, we quickly identified a host that first stop sending data to the peers, and confirmed that one of the NICs in the host was the culprit.

With the insights provided by the Fault Analyzer, the operations team was able to quickly isolate the faulty host, preventing further job failures. The root cause was confirmed as a NIC malfunction, and the training job was rescheduled on healthy nodes, minimizing downtime and resource loss.

\textbf{Model Code Issue Triage} As model parallelism patterns become increasingly complex, model developers frequently introduce bugs during development. Traditionally, when such bugs occurred, they often manifested as NCCL timeouts in PyTorch, providing little information about the underlying cause. Developers would then need to manually add logs, rerun jobs, and painstakingly parse outputs to identify the issue, making the debugging process both challenging and time-consuming.

A developer encountered a collective operation timeout during distributed training. The error surfaced as a generic NCCL timeout, with no indication of which collective operation or host was responsible, significantly hindering rapid diagnosis.

The Fault Analyzer automatically detects inter-collective dependencies to identify the specific collective operation likely causing the timeout. In this case, we identified that the issue was caused by a rank not joining the last collective in the Tensor Parallelism Process Group.
The analyzer then provides real-time status for each collective kernel (not running, running, finished) on every host. In this case, we identified that all the other ranks in the process group had entered the collective kernel, indicating that the missing rank was likely the culprit.

With the insights from the Fault Analyzer, the developer checked the model code and confirmed that the issue was caused by a bug that made that rank hang before scheduling the latest collective in the TP PG. The Fault Analyzer eliminated the need for extensive manual logging and log parsing, drastically improving the efficiency of triaging issues for developers.

\subsection{Performance Observability}

Meta’s large-scale AI workloads run across thousands of GPUs and servers, relying on RDMA for collective communication (e.g., AllReduce, AllGather, AllToAll). At this scale, even small inefficiencies or bottlenecks in the network stack can have a massive impact on overall job performance. Traditional profiling tools often lack the granularity or context to pinpoint issues at the transport layer, especially for collective operations.

The \textbf{Perf profiler} is a profiling framework in NCCLX designed to monitor and analyze IB transport-level events, such as RDMA Work Queue Element (WQE) events and control messages. CtranProfiler has three main modules:

\begin{enumerate}
    \item \textbf{AlgoProfiler}: Collects timestamps throughout the execution of a collective operation to break down and measure latency for different stages (buffer registration, control message synchronization, data transfer). Helps pinpoint which stage is the bottleneck in a slow collective, e.g., is it slow because of memory registration, network congestion, or control message delays.
    \item \textbf{SlowRankDetector}: Monitors per-rank network efficiency and identifies slow ranks. It reads WQE completion events generated from the CTran transport profiler and measures the bus bandwidth of each rank over a rolling window. A WQE completion event gives us information such as the number of bytes sent and the post and completion timestamps of the WQE.
    \item \textbf{QueuePairProfiler}: Provides per-queue-pair (QP) performance metrics, such as idle time and post frequency. Tracks WQE events and exports per-queue performance traces for further analysis. Supports development and tuning of advanced load-balancing algorithms (like DQPLB), and helps correlate performance drops with specific QPs or network paths.
\end{enumerate}

The Perf profiler framework in NCCLX is designed to provide granular visibility into the performance of collective communication operations by instrumenting the CTran code to collect events at checkpoints throughout the data flow. Specifically, it captures events whenever RDMA work requests are posted or completed, and when control or data messages are sent or received between ranks. To ensure accurate cross-rank correlation, all events are timestamped using Precision Time Protocol (PTP), which synchronizes clocks across distributed hosts.

Once collected, these events are processed by specialized modules---such as AlgoProfiler, SlowRankDetector, and QueuePairProfiler---which are enabled via environment variables and subscribe only to the event types relevant to their analysis. Each module implements custom logic to aggregate, analyze, and export the data in real time or for post-mortem investigation.

Focusing on the AlgoProfiler module, it breaks down the execution of collective operations into distinct stages, collecting detailed timing information for buffer registration (the process of preparing memory for RDMA transfers), control message synchronization (coordinating the start and completion of operations across ranks), and the actual data transfer phase. Every profiling record generated is tagged with metadata including the job name, device, group, message size, and iteration number, which allows engineers to filter and zoom in on specific hosts, steps, or collective operations when investigating performance issues. This comprehensive approach enables targeted root cause analysis, optimization, and monitoring of distributed training jobs at Meta’s scale.

\begin{figure}[tp]
        \centering
        \includegraphics[width=0.5\textwidth]{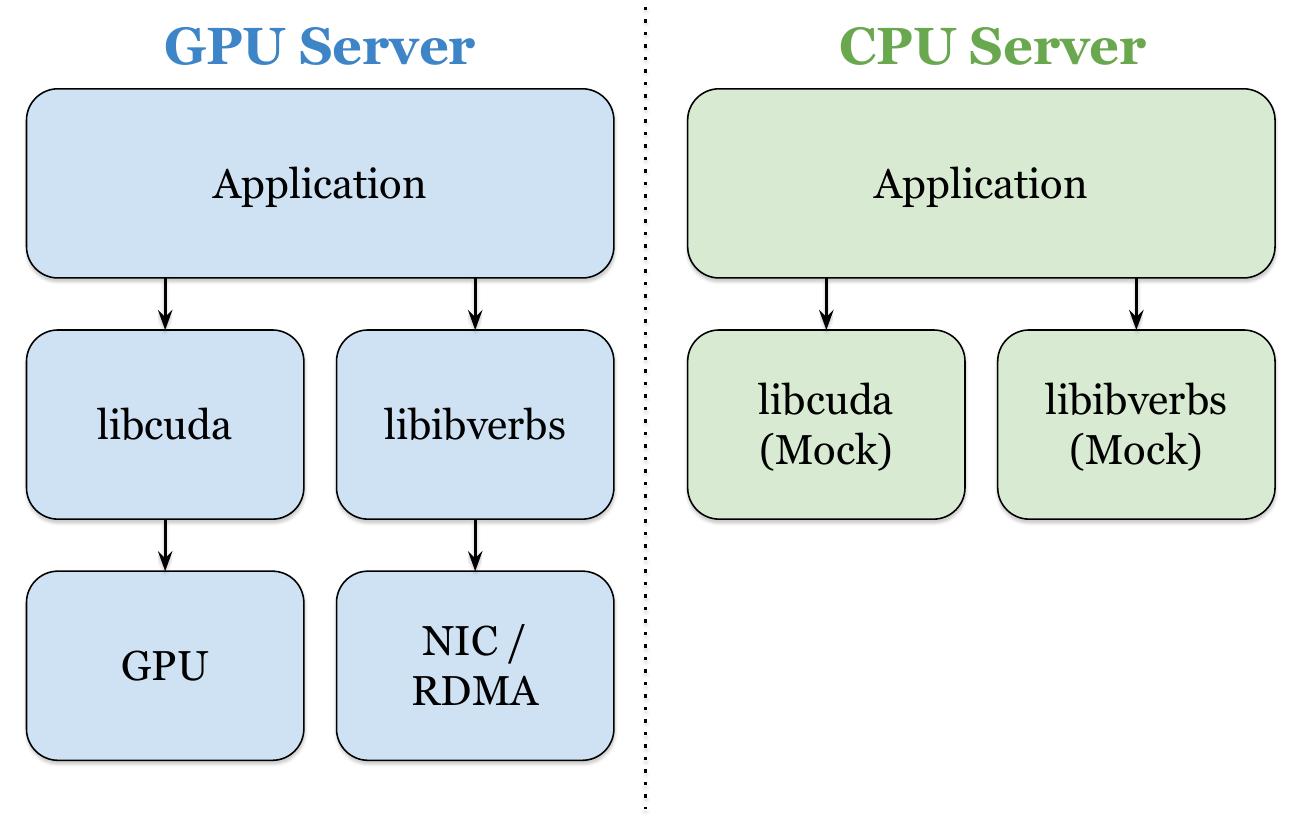}
        \caption{Comparison between NCCL program on GPU and CPU.}
        \label{fig:cpu-emulation}
\end{figure}

\subsection{CPU emulation}

To enable efficient development at the scale of 100K GPUs and beyond, novel testing methodologies are essential for validation without incurring massive resource requirements. We built a CPU emulation framework that allows large-scale testing—at 100K+ scale—without the need for extensive GPU resources.  This is critical for testing NCCL initialization and reproducing failure scenarios  at scale. Our CPU emulation framework mocks CUDA and RDMA libraries through custom shared libraries loaded via \texttt{LD\_LIBRARY\_PATH} manipulation, as shown in \autoref{fig:cpu-emulation}. This enables running unmodified NCCL code on CPU clusters. Further packing 32 processes per host can enable 96K scale testing with only 3,072 CPU servers. CPU emulation revealed critical bottlenecks invisible at smaller scales - notably busy-loop on I/O and $O(N^2)$ implementation complexity.

For example, we use CPU emulation to accurately measure the initialization at scale, identify key bottlenecks, and validate the effectiveness of the solutions with confidence. In addition to performance, we also discovered system limitation where scaling beyond 64K ranks. Specifically, we figured that TCP connections were silently getting dropped due to System's Listen Queue size overflow which at 64K scale. To address this, we implemented re-connection retries with exponential back-off as a solution.

%% file: related.tex
\section{Related Works}
\subsection{Traditional Communication Libraries for HPC and AI}
Several communication libraries have been developed to support large-scale parallel computing in both traditional High-Performance Computing (HPC) and modern AI workloads. The Message Passing Interface (MPI) is a classical host-driven library that offers rich semantics for point-to-point and collective communications \citep{mpi}. Its host-driven nature, however, can introduce overheads for GPU-to-GPU communication compared to more modern, device-centric approaches. A prominent example of a device-centric library is the NVIDIA Collective Communications Library (NCCL), a widely-used, kernel-driven, and copy-based library for intra-GPU and inter-GPU communication \citep{nccl}. Building on this, the Microsoft Collective Communication Library (MSCCL) provides a platform for custom collective algorithms through a domain-specific language, allowing for significant performance gains over standard NCCL for specific topologies and message sizes \citep{msccl}. Similarly, the Topology Aware Collective Communication Library (TACCL) synthesizes optimal collective algorithms for a given hardware configuration using user-provided "communication sketches" \citep{taccl}. These generated algorithms can then be registered and executed within the MSCCL framework. Another library, NVSHMEM, provides device-initiated one-sided communication and collectives, enabling direct GPU-to-GPU data transfer without CPU involvement, though its flexibility is limited by symmetric memory semantics \citep{nvshmem}. The Ultra and Unified CCL (UCCL) is a software transport layer that decouples data and control paths to provide an extensible, high-performance, and vendor-agnostic solution for GPU networking \citep{uccl}. UCCL focuses on transport innovations to improve performance for various ML workloads and can serve as a drop-in replacement for NCCL/RCCL with notable speedups.

\subsection{Communication Libraries for DL/LLM}
The specific requirements of Deep Learning  and LLMs have led to the development of specialized communication libraries. DeepEP, developed by DeepSeek AI, is a specialized communication library for Mixture-of-Experts (MoE) models, providing high-throughput, low-latency all-to-all GPU kernels for dispatch and combine operations \citep{deepep}. It leverages NVSHMEM for efficient one-sided communication and is designed to overlap communication and computation without occupying precious Streaming Multiprocessor  resources. Similarly, ByteDance's MegaScale is a full-stack production system for training LLMs at an unprecedented scale \citep{megascale}. Its MegaScale-MoE component specifically addresses communication bottlenecks in MoE models, demonstrating significant efficiency improvements.

Recent innovations have also focused on fine-grained optimization. For example, Flux is a communication-overlapping library for dense and MoE models that achieves fast, software-based communication overlap by over-decomposing communication and computation into fine-grained operations and fusing them into larger kernels \citep{flux}. This approach can effectively hide a significant portion of communication latency. Another advancement is MetaShuffling, a new MoE inference solution that efficiently deploys Llama4 models by directly sorting tokens based on their routed expert ID \citep{metashuffling}. This mechanism avoids the padding and slicing overheads of traditional methods, which can lead to increased memory usage and reduced kernel efficiency.

While many libraries are kernel-driven and copy-based, this work introduces a custom transport (CTran) stack that operates on a zero-copy, host-driven framework. This approach enables RDMA to be issued directly from the user source buffer to the destination buffer, which in turn reduces GPU resource contention, including HBM bandwidth and SMs, by eliminating unnecessary device-to-device copies. For inference workloads, a low-latency, host-driven AllToAll implementation is detailed, which is notable for its performance without requiring a separate communication runtime. The library also addresses the operational challenges of massive-scale systems by incorporating fault-tolerant collectives, allowing training to continue through hardware failures, and an automatic failure analyzer that significantly reduces debugging time. By minimizing GPU resource overhead and offering a GPU-resident collective scheme, which avoids data padding, this work provides a robust and efficient solution for both training and inference in production environments.

%% file: conclusion.tex
\section{Conclusion}
This paper has demonstrated that traditional collective communication methods are insufficient for the scale of modern LLMs, particularly for workloads on clusters of 100,000 GPUs and beyond. We have presented the NCCLX framework, a novel, practically-designed system from Meta that addresses these limitations by providing a unified solution for both synchronous training and low-latency inference. The framework's success, evidenced by its superior communication efficiency during the evaluation with the Llama4 model and the implementation of algorithms like CTran, confirms its role in enabling large-scale distributed machine learning. The robust solution provided by NCCLX not only resolves current communication bottlenecks but also paves the way for the next generation of models that will demand even greater scale and efficiency. This work underscores the importance of co-designing communication infrastructure with the computational needs of cutting-edge AI, ensuring that software and hardware advancements proceed in tandem to unlock the full potential of large-scale AI research and deployment.

%% file: acknowledgements.tex
\section*{Acknowledgements}
Many current and former people at Meta have contributed to productionizing collective communication libraries for AI training and inference over the years. Besides authors, the following people has made significant contribution to the NCCLX and CTran: Ajay Hotchandani, Kai Luo, Deep Shah, and Lukasz Wesolowski.

This work is a close collaboration with our partners in Meta's AI teams. We would like to express our gratitude to the following individuals for their invaluable help and contributions: Kunal Bhalla, Sai Jayesh Bondu, Will Constable, Zachary DeVito, Mikel Jimenez Fernandez, Wenyin Fu, Naman Goyal, Andrew Gu, Yuchen Hao, Eliot Hedeman, Jianyu Huang, Jenya Lee,  Shikai Li, Keyu Man, Mustafa Ozdal, Jason Park, Jongsoo Park,  Shangfu Peng, Sarunya Pumma, Tristan Rice, Ahmed Sharif, Omkar Salpekar, Zoey Sun, Michael Suo, Rohan Varma, Shawn Xu, Vedanuj Goswami, Weiwei Chu, Wei Sun, Jie Wang, Xiaodong Wang, Yifu Wang, Ke Wen, Ji Yang, Jiecao Yu, and Haoci Zhang.

We are also grateful for the project support and guidance from Balaji Balasubramanian, Omar Baldonado, Harish Kumar Chandrappa, Shashi Gandham, Guna Lakshminarayanan, Teng Li, Maxim Naumov, Mathew Oldham, Chunqiang Tang, Srinivas Vaidyanathan, and Chuanhao Zhuge.

We would like to extend our sincere thanks to our partners in the NVIDIA NCCL team for their insightful discussions and technical support.